\documentclass[twocolumn]{aastex631}

\usepackage{xspace}
\usepackage{CJK}
\usepackage{rotating}
\usepackage{afterpage}
\usepackage{changepage}
\usepackage{sidecap}
\usepackage{comment}

\hypersetup{
    colorlinks=true,
    linkcolor=blue,
    filecolor=magenta,      
    urlcolor=cyan,
}

\usepackage[normalem]{ulem} % for regular text; use \sout{}
\usepackage{cancel} % for equations; use \cancel{}

\usepackage{color}

\newcommand{\Mstar}{\ensuremath{M_{\rm{*}}}\xspace}

\newcommand{\Msol}{\ensuremath{M_\odot}\xspace}

\newcommand{\oiii}{[O{\scriptsize III}]\xspace}

\newcommand{\squiggle}{SQuIGG\ensuremath{\vec{L}}E\xspace}
\newcommand{\PSBI}{SDSS\,J1448+1010\xspace}
\newcommand{\PSBII}{SDSS\,J2258+2313\xspace}

\newcommand{\oiiihb}{\ensuremath{[\mathrm{O\,III}]/\mathrm{H}\beta}\xspace}

\shortauthors{D'Onofrio, et~al.}
\shorttitle{Radio AGN in \squiggle PSBs}

\begin{document}
\begin{CJK*}{UTF8}{gbsn}

\title{\squiggle: Radio AGN Found Predominantly in Young, Gas-rich Post-starburst Galaxies at $z\sim0.7$}

\correspondingauthor{Vincenzo~R.~D'Onofrio}
\email{donofr19@tamu.edu}

\author[0000-0002-1759-6205]{Vincenzo~R.~D'Onofrio}
\affiliation{Department of Physics and Astronomy and George P. and Cynthia Woods Mitchell Institute for Fundamental Physics and Astronomy, Texas A\&M University, 4242 TAMU, College Station, TX 77843-4242, US}

\author[0000-0003-3256-5615]{Justin~S.~Spilker}
\affiliation{Department of Physics and Astronomy and George P. and Cynthia Woods Mitchell Institute for Fundamental Physics and Astronomy, Texas A\&M University, 4242 TAMU, College Station, TX 77843-4242, US}

\author[0000-0001-5785-7038]{Krista~L.~Smith}
\affiliation{Department of Physics and Astronomy and George P. and Cynthia Woods Mitchell Institute for Fundamental Physics and Astronomy, Texas A\&M University, 4242 TAMU, College Station, TX 77843-4242, US}

\author[0000-0001-5063-8254]{Rachel~Bezanson}
\affiliation{Department of Physics and Astronomy and PITT PACC, University of Pittsburgh, Pittsburgh, PA 15260, USA}

\author[0000-0002-1109-1919]{Robert~Feldmann}
\affiliation{Department of Astrophysics, Universit\"at Z\"urich, Winterthurerstrasse 190, Zurich CH-8057, Switzerland}

\author[0009-0009-9810-3334]{Alex~J.~Geiger}
\affiliation{Department of Physics and Astronomy and George P. and Cynthia Woods Mitchell Institute for Fundamental Physics and Astronomy, Texas A\&M University, 4242 TAMU, College Station, TX 77843-4242, US}

\author[0000-0003-4700-663X]{Andy~D.~Goulding}
\affiliation{Department of Astrophysical Sciences, Princeton University, 4 Ivy Lane, Princeton, NJ 08544, USA}

\author[0000-0002-5612-3427]{Jenny~E.~Greene}
\affiliation{Department of Astrophysical Sciences, Princeton University, 4 Ivy Lane, Princeton, NJ 08544, USA}

\author[0000-0002-7613-9872]{Mariska~Kriek}
\affiliation{Leiden Observatory, Leiden University, P.O. Box 9513, 2300 RA Leiden, The Netherlands}

\author[0009-0005-4226-0964]{Anika~Kumar}
\affiliation{Laboratory for Multiwavelength Astrophysics, School of Physics and Astronomy, Rochester Institute of Technology, 84 Lomb Memorial Drive, Rochester, NY 14623, USA}

\author[0000-0002-0696-6952]{Yuanze~Luo}
\affiliation{Department of Physics and Astronomy and George P. and Cynthia Woods Mitchell Institute for Fundamental Physics and Astronomy, Texas A\&M University, 4242 TAMU, College Station, TX 77843-4242, US}

\author[0000-0002-7064-4309]{Desika~Narayanan}
\affiliation{Department of Astronomy, University of Florida, 211 Bryant Space Science Center, Gainesville, FL 32611, USA}

\author[0000-0003-4075-7393]{David~J.~Setton}
\altaffiliation{NHFP Hubble Fellow}
\affiliation{The William H. Miller III Department of Physics and Astronomy, 3400 N. Charles Street, Baltimore, MD 21218}

\author[0000-0002-1714-1905]{Katherine~A.~Suess}
\affiliation{Department for Astrophysical \& Planetary Science, University of Colorado, Boulder, CO 80309, USA}

\author[0000-0003-1535-4277]{Margaret~E.~Verrico}
\affiliation{University of Illinois Urbana-Champaign Department of Astronomy, University of Illinois, 1002 W. Green St., Urbana, IL 61801, USA}
\affiliation{Center for AstroPhysical Surveys, National Center for Supercomputing Applications, 1205 West Clark Street, Urbana, IL 61801, USA}

\author[0000-0002-6768-8335]{Pengpei~Zhu}
\affiliation{Cosmic Dawn Center (DAWN), Denmark}
\affiliation{DTU Space, Technical University of Denmark, Elektrovej 327, 2800 Kgs. Lyngby, Denmark}

%%%%%%%%%%%%%%%%%%%%%%%%%%%%%%%%%%%%%%%%%%%%%%%%%%%%%%%%%%%%%%%%%%%%%%%%%%%%%%%%%%%%%
%%%%%%%%%%%%%%%%%%%%%%%%%%%%%%%%%%%% ABSTRACT %%%%%%%%%%%%%%%%%%%%%%%%%%%%%%%%%%%%%%%
%%%%%%%%%%%%%%%%%%%%%%%%%%%%%%%%%%%%%%%%%%%%%%%%%%%%%%%%%%%%%%%%%%%%%%%%%%%%%%%%%%%%%
\begin{abstract}
The processes that rapidly quench massive galaxies and maintain their quiescence remain poorly understood. Substantial molecular gas reservoirs can survive this shutdown in post-starburst galaxies, raising the question of what prevents the remaining gas from efficiently forming stars. We present deep, subarcsecond VLA 6\,GHz observations of 52 massive \squiggle post-starburst galaxies at $z\sim0.7$ to test the prevalence and mechanical impact of radio activity during the gas-rich post-burst phase. We detect 6\,GHz continuum emission in 18/52 galaxies ($\approx$35\%), spanning $\log(L_{1.4\,{\rm GHz}}/{\rm W\,Hz^{-1}})\simeq23.0$--$25.0$ and generally exceeding the contributions expected from ongoing star formation and the fading starburst, favoring an AGN origin. Ancillary data at other wavelengths provide little evidence for a population of luminous, radiatively efficient AGN, suggesting that these sources are instead dominated by comparatively weak, mechanically powered radio AGN. Radio activity is strongly concentrated within the first $\sim$200\,Myr post-burst and is predominantly associated with galaxies retaining substantial gas reservoirs. Above a common radio luminosity threshold, the radio detection fraction is enhanced by a factor of $\approx$2.5 relative to the radio AGN candidate fraction in mass- and redshift-matched quiescent LEGA-C galaxies, primarily due to lower power sources. For efficient jet--ISM coupling, the inferred mechanical power could supply the approximate energy required to unbind the molecular gas on timescales comparable to its observed decline. Together, these results identify a population of radiatively inefficient AGN hosted predominantly by young, gas-rich post-starburst galaxies at $z\sim0.7$ and suggest that they may have a role in maintaining inefficient star formation.   
\end{abstract}

%%%%%%%%%%%%%%%%%%%%%%%%%%%%%%%%%%%%%%%%%%%%%%%%%%%%%%%%%%%%%%%%%%%%%%%%%%%%%%%%%%%%%
%%%%%%%%%%%%%%%%%%%%%%%%%%%%%%%%%% Introduction %%%%%%%%%%%%%%%%%%%%%%%%%%%%%%%%%%%%%
%%%%%%%%%%%%%%%%%%%%%%%%%%%%%%%%%%%%%%%%%%%%%%%%%%%%%%%%%%%%%%%%%%%%%%%%%%%%%%%%%%%%%
\section{Introduction}
\label{intro}

Understanding how massive galaxies rapidly stop forming stars and subsequently remain quiescent is a central unresolved problem in galaxy evolution (see \citealt{whitakerQuenchingStarFormation2026} for a recent review). Feedback from active galactic nuclei (AGN) is frequently invoked to regulate this transition by expelling or dynamically perturbing the interstellar medium (ISM) \citep[e.g.,][]{springelModellingFeedbackStars2005,dimatteoEnergyInputQuasars2005,hopkinsUnifiedMergerdrivenModel2006,crotonManyLivesActive2006,fabianObservationalEvidenceAGN2012,heckmanCoEvolutionGalaxiesSupermassive2014,veilleuxCoolOutflowsGalaxies2020}. Establishing this connection observationally is difficult because black hole accretion varies on timescales much shorter than those associated with galaxy wide quenching. An AGN observed after star formation has ceased may or may not have been active during the shutdown, while the absence of a luminous AGN does not exclude earlier activity or a weak, recurrent source capable of affecting the remaining gas \citep[e.g.,][]{wildTimingAGNStarbursts2010,yesufStarburstQuiescenceTesting2014,frenchAGNLightEchoes2023}. Determining the role of AGN feedback therefore requires systems with well constrained recent evolutionary histories.

Post-starburst galaxies (PSBs) provide a particularly direct view of this transition. Their strong Balmer absorption and weak nebular emission indicate that a substantial episode of star formation ended rapidly within the previous $\sim$0.1--1\,Gyr, providing an approximate clock for the time elapsed since the burst \citep[e.g.,][]{zabludoffEnvironmentE+AGalaxies1996,wildEvolutionPoststarburstGalaxies2016,pawlikOriginsPoststarburstGalaxies2018,frenchEvolutionPoststarburstPhase2021}. Importantly, many PSBs retain substantial molecular gas reservoirs despite their sharply reduced star formation rates (SFRs), demonstrating that the cessation of star formation does not necessarily require the immediate removal of the cold gas \citep[e.g.,][]{frenchDiscoveryLargeMolecular2015a,rowlandsEvolutionColdInterstellar2015a,alataloShockedPOststarburstGalaxy2016a,suessMassiveQuenchedGalaxies2017a,bezansonNowYouSee2022,settonSQuIGGBuriedStar2025c}. Instead, the surviving gas may remain inefficient at forming stars because it is too diffuse, turbulent, dynamically disturbed, or unable to cool into a dense star forming phase \citep[e.g.,][]{frenchWhyPoststarburstGalaxies2018,smercinaFallResolvingMolecular2022b,frenchStateMolecularGas2023c,donofrioQuenchingTidalGas2025a,otterClumpyDenseGas2026a}. PSBs therefore offer a uniquely short time period over which to test whether AGN activity accompanies the rapid decline in star formation or acts later upon the residual gas to maintain quiescence.

The Studying Quenching in Intermediate-$z$ Galaxies: Gas, angu$\vec{L}$ar momentum, and Evolution (\squiggle) survey investigates massive PSBs at $z\sim0.7$ \citep{suessTextbackslashmathrmSQuIGGecLEStudyingQuenching2022}. Atacama Large Millimeter/submillimeter Array (ALMA) observations show that many of the youngest \squiggle galaxies retain large molecular gas reservoirs immediately after the rapid suppression of star formation, followed by a rapid decline in their gas content over the subsequent $\sim$100--200\,Myr \citep{bezansonNowYouSee2022,settonSQuIGGBuriedStar2025c}. In individual systems a substantial fraction of this gas has been redistributed into extended tidal structures, while the material remaining in the central galaxy is turbulent and inefficient at forming stars \citep{spilkerStarFormationSuppression2022,donofrioQuenchingTidalGas2025a}. Mergers have been commonly invoked to contribute to the initial suppression of star formation \citep[e.g.,][]{toomreGalacticBridgesTails1972,hibbardNeutralHydrogenDistribution2000,spilkerStarFormationSuppression2022,donofrioQuenchingTidalGas2025a}, but they do not fully explain why the gas retained remains quiescent.

AGN feedback is commonly separated into a radiatively efficient `quasar mode' in which luminous accretion can drive winds that heat or expel gas, and a radiatively inefficient `radio mode' where a larger fraction of the accretion power emerges mechanically through jets \citep[e.g.,][]{crotonManyLivesActive2006,fabianObservationalEvidenceAGN2012,heckmanCoEvolutionGalaxiesSupermassive2014,hardcastleCrostonRadioGalaxiesFeedback2020}. The former may contribute to the rapid shutdown of a gas-rich starburst, whereas mechanical energy input from low luminosity AGN provides one possible mechanism for maintaining the suppressed state \citep[e.g.,][]{crotonManyLivesActive2006,heckmanCoEvolutionGalaxiesSupermassive2014,hardcastleCrostonRadioGalaxiesFeedback2020}. Evidence for enhanced AGN activity during the PSB phase has nevertheless remained mixed \citep[e.g.,][]{yesufStarburstQuiescenceTesting2014,frenchWhyPoststarburstGalaxies2018,frenchEvolutionPoststarburstPhase2021,luoMultiwavelengthEvaluationAGN2026}. Optical studies find AGN preferentially among some of the youngest PSBs, although this association may reflect the availability of gas for black hole accretion rather than a causal connection to quenching \citep[e.g.,][]{wildTimingAGNStarbursts2010,greeneRoleActiveGalactic2020,frenchAGNLightEchoes2023}. X-ray and mid-infrared studies similarly indicate that luminous radiatively efficient AGN are not ubiquitous among PSBs and may be visible for only a small fraction of the transition \citep[e.g.,][]{almainiNoEvidenceExcessAGN2025,luoMultiwavelengthEvaluationAGN2026}. Radio observations offer a complementary probe because synchrotron emission is insensitive to dust obscuration and can reveal mechanically dominated, radiatively inefficient accretion that is weak in conventional optical, infrared, and X-ray diagnostics \citep[e.g.,][]{condonRadioEmissionNormal1992,smolcicCosmicEvolutionRadio2009,heckmanCoEvolutionGalaxiesSupermassive2014,hardcastleCrostonRadioGalaxiesFeedback2020}. These selection methods therefore probe partly distinct AGN populations, with radio selected sources preferentially occupying lower Eddington ratios than X-ray and infrared selected AGN \citep{hickoxHostGalaxiesClustering2009}. Radio studies support a picture in which compact or low power radio sources occur in some massive PSBs, but their inferred incidence depends strongly on stellar mass, luminosity threshold, and survey depth \citep[e.g.,][]{shinRadioEmissionActive2011,greeneRoleActiveGalactic2020,luoMultiwavelengthEvaluationAGN2026,patilTracingRadioAGNDriven2026}. Previous radio studies of \squiggle were limited primarily to the comparatively shallow observations from the Faint Images of the Radio Sky at Twenty-cm (FIRST) survey and therefore selected only the most radio luminous systems \citep{greeneRoleActiveGalactic2020}. These observations established that radio luminous AGN occur among massive, recently quenched galaxies but could not test whether lower power radio activity is common across the broader population. This distinction is important if AGN feedback operates through recurrent, moderate power episodes rather than a single luminous event \citep[e.g.,][]{bestHostGalaxiesRadioloud2005,sabaterLoTSSViewRadio2019,hardcastleCrostonRadioGalaxiesFeedback2020}. A sensitive radio census is therefore needed to establish how common low luminosity radio activity is in PSBs and to characterize its possible connection to PSB evolution.

In this work, we present deep Very Large Array (VLA) $C$-band (6\,GHz) imaging of 52 massive \squiggle PSBs at $z\sim0.7$ to characterize faint radio activity during the early post-burst phase and assess whether its mechanical power could affect the surviving molecular gas. Section~\ref{data} describes the sample and observations. Section~\ref{analysis} presents the radio measurements and tests of their origin. In Section~\ref{Results} we examine their connection to galaxy properties, comparison populations, and the maintenance of low SFRs. We summarize our conclusions in Section~\ref{Conclusions}. Throughout this work, we assume a \citet{chabrierGalacticStellarSubstellar2003} initial mass function and a flat $\Lambda$CDM cosmology with $\Omega_{\rm m}=0.3$ and $h=0.7$.

%%%%%%%%%%%%%%%%%%%%%%%%%%%%%%%%%%%%%%%%%%%%%%%%%%%%%%%%%%%%%%%%%%%%%%%%%%%%%%%%%%%%%
%%%%%%%%%%%%%%%%%%%%%%%%%%%%%%%%% Data & Methods %%%%%%%%%%%%%%%%%%%%%%%%%%%%%%%%%%%%
%%%%%%%%%%%%%%%%%%%%%%%%%%%%%%%%%%%%%%%%%%%%%%%%%%%%%%%%%%%%%%%%%%%%%%%%%%%%%%%%%%%%%
\section{Data and Methods} \label{data}

\subsection{Overview of the Presented Sample}
\label{label:targets}

\begin{figure}[t]
    \centering
    \includegraphics[width=0.95\columnwidth]
    {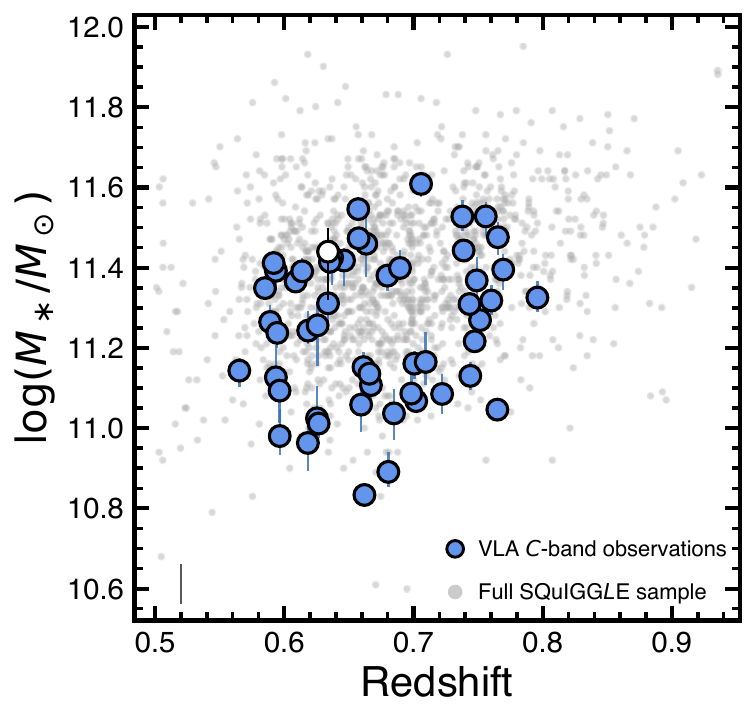}
    \caption{Stellar mass versus redshift for the \squiggle parent catalog (gray) and the 52 galaxies targeted by our VLA $C$-band observations (blue). Of the 52 VLA targets, all but one have ALMA CO(2--1) measurements (SDSS~J1235+0246, white). The median stellar mass uncertainty is shown for the full \squiggle sample (black bar). The VLA sample spans $0.6\lesssim z \lesssim 0.8$ and $10.8\lesssim\log(M_\ast/M_\odot)\lesssim11.6$, providing broad coverage of the full \squiggle population.}
    \label{fig:Mstar_redshift}
\end{figure}

The 52 massive \squiggle PSBs ($\log\,(\Mstar/\Msol) \approx 10.8-11.6$) in this work are drawn from the parent sample of 1318 PSBs at $z\approx0.5-0.9$ \citep{suessTextbackslashmathrmSQuIGGecLEStudyingQuenching2022}. Figure \ref{fig:Mstar_redshift} shows the stellar mass versus redshift distribution of the full \squiggle sample along with the targets with VLA $C$-band measurements. The VLA sample was designed to overlap nearly completely with the existing ALMA CO(2--1) follow-up \citep{suessMassiveQuenchedGalaxies2017a,bezansonNowYouSee2022,settonSQuIGGBuriedStar2025c}; all but one object have published CO(2--1) measurements, with SDSS\,J1235+0246 as the single exception. The expanded ALMA program presented in \citet{settonSQuIGGBuriedStar2025c} was constructed using a median-distance Monte Carlo selection in order to broaden the earlier follow-up presented in \citet{bezansonNowYouSee2022} which was preferentially weighted toward higher stellar masses and lower redshifts, to provide substantially more representative coverage of the full \squiggle population (Figure \ref{fig:Mstar_redshift}). For the overview of the sample we adopt the spectroscopic redshifts and stellar masses from \citet{suessTextbackslashmathrmSQuIGGecLEStudyingQuenching2022}. The combination of VLA and ALMA coverage allows us to test whether radio activity emerges while substantial molecular reservoirs are still present and whether it's connected to the rapid decline of cold gas.

\subsection{\squiggle Properties}
\label{label:squiggleprops}

\begin{table*}[ht!]
\centering
\caption{VLA and Ancillary \squiggle Source Measurements.}
\label{tab:vla_props}
\begingroup
\scriptsize
\setlength{\tabcolsep}{3pt}
\renewcommand{\arraystretch}{1.0}
\begin{tabular*}{0.99\textwidth}{@{\extracolsep{\fill}}lccccccc@{}}
\hline\hline
Object & $z$ & $M_{\mathrm{H}_2}$\tablenotemark{a} &
$\sigma_\star$\tablenotemark{b} &
$S_{\rm int,6\,\mathrm{GHz}}$ &
$S_{\rm peak,6\,\mathrm{GHz}}$ &
$D$ & $L_{1.4\,\mathrm{GHz}}$ \\
& & [$10^{10}\,M_\odot$] & [km s$^{-1}$] &
[$10^{1}\,\mu$Jy] &
[$10^{1}\,\mu$Jy beam$^{-1}$] &
[kpc] & [$10^{22}\,\mathrm{W\,Hz^{-1}}$] \\
\hline
J0027+0129 & 0.5851 & $<0.24$ & $179 \pm 40$ & $<1.38$ & -- & -- & $<5.36$ \\
J0044-0054 & 0.7653 & $<0.71$ & $122 \pm 51$ & $9.30 \pm 0.46$ & $9.61 \pm 0.45$ & $<0.26$ & $69.6 \pm 3.44$ \\
J0046-0147 & 0.6088 & $<0.24$ & $196 \pm 38$ & $<1.47$ & -- & -- & $<6.29$ \\
J0131+0034 & 0.6183 & $0.11 \pm 0.03$ & $197 \pm 51$ & $<1.26$ & -- & -- & $<5.60$ \\
J0153-0207 & 0.6658 & $0.95 \pm 0.09$ & $186 \pm 31$ & $<1.23$ & -- & -- & $<6.55$ \\
J0221-0646 & 0.6613 & $<0.21$ & $184 \pm 77$ & $<1.29$ & -- & -- & $<6.75$ \\
J0224-0630 & 0.7515 & $2.00 \pm 0.20$ & $204 \pm 82$ & $<1.35$ & -- & -- & $<9.66$ \\
J0224-0034 & 0.7433 & $1.59 \pm 0.18$ & $209 \pm 70$ & $208 \pm 0.43$ & $105 \pm 0.46$ & $3.83 \pm 0.03$ & $1450 \pm 3.00$ \\
J0224+0015 & 0.6849 & $<0.31$ & $239 \pm 39$ & $<1.26$ & -- & -- & $<7.18$ \\
J0227-0225 & 0.6139 & $2.32 \pm 0.09$ & $237 \pm 32$ & $4.49 \pm 0.47$ & $1.41 \pm 0.36$ & $10.13 \pm 3.50$ & $19.6 \pm 2.05$ \\
J0232-0331 & 0.7388 & $<0.43$ & $200 \pm 66$ & $<1.41$ & -- & -- & $<9.68$ \\
J0233+0052 & 0.5918 & $<0.22$ & $213 \pm 40$ & $<1.41$ & -- & -- & $<5.63$ \\
J0237+0123 & 0.7094 & $1.75 \pm 0.17$ & $153 \pm 60$ & $1.45 \pm 0.48$ & $2.34 \pm 0.42$ & $<0.99$ & $9.01 \pm 2.98$ \\
J0240-0148 & 0.6895 & $<0.67$ & $193 \pm 19$ & $<1.47$ & -- & -- & $<8.52$ \\
J0753+2403 & 0.5652 & $<0.17$ & $196 \pm 19$ & $130 \pm 0.45$ & $127 \pm 0.45$ & $0.47 \pm 0.05$ & $464 \pm 1.61$ \\
J0851+0244 & 0.6595 & $1.51 \pm 0.16$ & $158 \pm 36$ & $<1.41$ & -- & -- & $<7.33$ \\
J0907+0423 & 0.6635 & $5.76 \pm 0.07$ & $169 \pm 45$ & $23.1 \pm 0.47$ & $22.9 \pm 0.47$ & $<0.09$ & $121 \pm 2.48$ \\
J0909-0108 & 0.7021 & $2.79 \pm 0.40$ & $136 \pm 31$ & $<1.38$ & -- & -- & $<8.36$ \\
J0910+0218 & 0.7694 & $6.11 \pm 0.15$ & $156 \pm 113$ & $3.02 \pm 0.45$ & $3.17 \pm 0.47$ & $<0.71$ & $22.7 \pm 3.41$ \\
J0912+1523 & 0.7473 & $3.39 \pm 0.14$ & $266 \pm 44$ & $<1.38$ & -- & -- & $<9.74$ \\
J0940-0008 & 0.6185 & $<0.44$ & $180 \pm 28$ & $<1.38$ & -- & -- & $<6.14$ \\
J1007+2330 & 0.6353 & $<0.34$ & $229 \pm 36$ & $<1.35$ & -- & -- & $<6.41$ \\
J1017-0003 & 0.5964 & $<0.26$ & $170 \pm 37$ & $11.0 \pm 0.45$ & $10.6 \pm 0.46$ & $0.86 \pm 0.37$ & $44.8 \pm 1.83$ \\
J1040+0223 & 0.6805 & $<0.42$ & $192 \pm 31$ & $<1.98$ & -- & -- & $<11.1$ \\
J1042+0500 & 0.6266 & $0.45 \pm 0.11$ & $151 \pm 28$ & $<2.07$ & -- & -- & $<9.50$ \\
J1046+0123 & 0.7649 & $<1.15$ & $179 \pm 53$ & $<2.16$ & -- & -- & $<16.1$ \\
J1053+2342 & 0.6370 & $<0.35$ & $187 \pm 21$ & $<1.35$ & -- & -- & $<6.45$ \\
J1109-0040 & 0.5935 & $1.54 \pm 0.12$ & $239 \pm 42$ & $<2.28$ & -- & -- & $<9.17$ \\
J1114+0115 & 0.7489 & $1.60 \pm 0.44$ & $170 \pm 75$ & $<2.34$ & -- & -- & $<16.6$ \\
J1141-0109 & 0.6575 & $<0.35$ & $247 \pm 24$ & $<2.16$ & -- & -- & $<11.1$ \\
J1142+0006 & 0.5935 & $4.49 \pm 0.10$ & $201 \pm 47$ & $13.0 \pm 0.71$ & $12.4 \pm 0.72$ & $<0.28$ & $52.3 \pm 2.86$ \\
J1148+0204 & 0.5965 & $<0.26$ & $199 \pm 22$ & $<1.35$ & -- & -- & $<5.50$ \\
J1157+0132 & 0.7559 & $11.30 \pm 0.16$ & $242 \pm 61$ & $9.69 \pm 0.44$ & $9.71 \pm 0.44$ & $<0.24$ & $70.4 \pm 3.19$ \\
J1203+1807 & 0.5946 & $<0.18$ & $235 \pm 37$ & $<1.38$ & -- & -- & $<5.57$ \\
J1211+0240 & 0.5890 & $0.47 \pm 0.11$ & $199 \pm 28$ & $1.64 \pm 0.45$ & $1.94 \pm 0.38$ & $<1.08$ & $6.32 \pm 1.78$ \\
J1222+0342 & 0.6798 & $0.78 \pm 0.21$ & $163 \pm 37$ & $<1.32$ & -- & -- & $<7.39$ \\
J1235+0246 & 0.6339 & -- & $164 \pm 44$ & $3.21 \pm 0.45$ & $3.56 \pm 0.45$ & $<0.60$ & $15.1 \pm 2.12$ \\
J1240-0057 & 0.7959 & $<1.34$ & $217 \pm 46$ & $<1.35$ & -- & -- & $<11.1$ \\
J1244+0248 & 0.6258 & $1.89 \pm 0.39$ & $153 \pm 67$ & $3.50 \pm 0.45$ & $2.94 \pm 0.49$ & $2.38 \pm 1.12$ & $16.0 \pm 2.06$ \\
J1332+0256 & 0.6983 & $0.64 \pm 0.16$ & $206 \pm 51$ & $<1.56$ & -- & -- & $<9.32$ \\
J1416+0255 & 0.7602 & $<0.43$ & $223 \pm 38$ & $<1.53$ & -- & -- & $<11.3$ \\
J1436+0447 & 0.6339 & $2.78 \pm 0.17$ & $171 \pm 51$ & $2.88 \pm 0.51$ & $3.17 \pm 0.48$ & $<0.71$ & $13.7 \pm 2.41$ \\
J1437+0311 & 0.6670 & $<0.35$ & $172 \pm 60$ & $<1.53$ & -- & -- & $<8.18$ \\
J1444-0006 & 0.7007 & $0.82 \pm 0.11$ & $264 \pm 79$ & $<1.53$ & -- & -- & $<9.22$ \\
J1448+1010 & 0.6462 & $1.93 \pm 0.12$ & $209 \pm 30$ & $10.2 \pm 0.39$ & $3.31 \pm 0.20$\tablenotemark{c} & $3.21 \pm 0.99$ & $50.5 \pm 1.93$ \\
J1449+0206 & 0.7440 & $0.75 \pm 0.20$ & $188 \pm 53$ & $3.00 \pm 0.53$ & $2.47 \pm 0.49$ & $<1.09$ & $20.9 \pm 3.70$ \\
J1455-0048 & 0.6253 & $<0.42$ & $141 \pm 35$ & $<1.59$ & -- & -- & $<7.26$ \\
J2202-0033 & 0.6573 & $0.65 \pm 0.08$ & $237 \pm 50$ & $10.5 \pm 0.52$ & $10.8 \pm 0.50$ & $<0.24$ & $54.1 \pm 2.68$ \\
J2213-0050 & 0.6620 & $<0.39$ & $131 \pm 34$ & $<1.56$ & -- & -- & $<8.19$ \\
J2232+0007 & 0.7222 & $<0.61$ & $183 \pm 68$ & $<1.56$ & -- & -- & $<10.1$ \\
J2258+2313 & 0.7058 & $5.24 \pm 0.22$ & $175 \pm 47$ & $1.58 \pm 0.66$ & $1.48 \pm 0.20$\tablenotemark{c} & $<1.10$\tablenotemark{c} & $9.70 \pm 4.05$ \\
J2310-0047 & 0.7378 & $3.48 \pm 0.14$ & $270 \pm 134$ & $<1.56$ & -- & -- & $<10.7$ \\
\hline
\end{tabular*}

\tablenotetext{a}{\raggedright $M_{\mathrm{H}_2}$ measurements for most objects are from \citet{bezansonNowYouSee2022} and \citet{settonSQuIGGBuriedStar2025c}. The measurements for the central galaxies of \PSBI\ and \PSBII\ come from \citet{spilkerStarFormationSuppression2022} and \citet{donofrioQuenchingTidalGas2025a}, respectively. These masses assume $r_{21}=1.0$ and $\alpha_{\rm CO}=4.0\,M_\odot\,({\rm K\,km\,s^{-1}\,pc^2})^{-1}$.}

\tablenotetext{b}{\raggedright Stellar velocity dispersions are from \citet{suessTextbackslashmathrmSQuIGGecLEStudyingQuenching2022}.}

\tablenotetext{c}{\raggedright From \citet{donofrioQuenchingTidalGas2025a}.}

\vspace*{6pt}
\small
\hspace*{-1.0em}%
\begin{minipage}[t]{0.98\linewidth}
\raggedright N\textsc{ote}---Upper limits are $3\sigma$, except for source diameter $D$ which is a $2\sigma$ upper limit following \citet{condon1997}.
\end{minipage}
\endgroup
\end{table*}

Throughout this work we combine the VLA measurements with various stellar population and molecular gas properties derived in previous \squiggle studies. We use the conventional measurements from the spectral energy distribution (SED) fitting in \citet{suessTextbackslashmathrmSQuIGGecLEStudyingQuenching2022} derived from the Sloan Digital Sky Survey (SDSS) optical spectroscopy and optical-near infrared photometry when concerning the full \squiggle sample (Figure \ref{fig:Mstar_redshift}), or when not included in the updated models in \citet{settonSQuIGGBuriedStar2025c} (SDSS~J0044$-$0054 and SDSS~J1235+0246). For galaxies in the expanded ALMA sample we adopt the panchromatic modeling of \citet{settonSQuIGGBuriedStar2025c}, which incorporates optical-to-far-infrared constraints and includes an independent star formation history (SFH) bin over the most recent 10\,Myr using two dust prescriptions: a `tied birth cloud' model in which the birth cloud and diffuse ISM attenuation are linked, and a `free birth cloud' model that permits additional optically obscured star formation. Pa$\alpha$ measurements of three young, CO-luminous \squiggle galaxies favor the higher obscured SFRs permitted by the free birth cloud models \citep{settonSpatiallyResolvedPaalpha2026}, although even these estimates generally remain insufficient to explain the observed radio luminosities (see Section~\ref{label:origin}). When a single fiducial value is required, we adopt the former unless otherwise stated. From \citet{settonSQuIGGBuriedStar2025c} we use the updated PSB age $t_{\rm PSB}$, which is defined as the lookback time at which 99\% of the stellar mass formed during the final 1\,Gyr before observation. Throughout this work, `younger' and `older' PSBs refer to shorter and longer times since the end of the recent burst, respectively, rather than the ages of their overall stellar populations. Additionally, we adopt the ALMA CO(2--1) measurements from \citet{bezansonNowYouSee2022} and \citet{settonSQuIGGBuriedStar2025c}, where molecular gas masses assume $r_{21}=1.0$ and $\alpha_{\rm CO}=4.0\,M_\odot\,({\rm K\,km\,s^{-1}\,pc^2})^{-1}$. We refer to the above works for full descriptions of the spectrophotometric modeling and measurement derivation. 

\begin{figure*}[ht!]
    \centering
    \includegraphics[width=\textwidth]
    {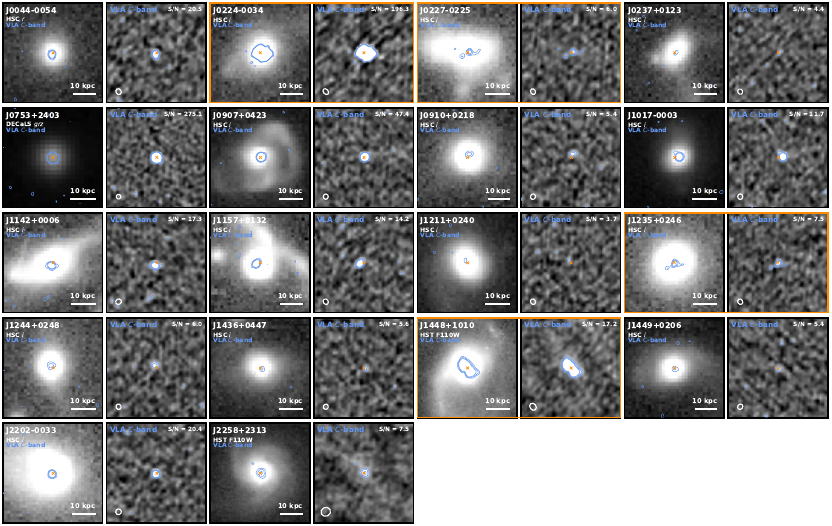}
    \caption{VLA $C$-band detections in the \squiggle sample. For each galaxy, the left panel shows the best available optical imaging with VLA 6\,GHz contours (3, 5, 7$\sigma$), while the right panel shows the VLA image with the same contours. The synthesized beam is shown in each radio panel. We use the best available optical or near-infrared imaging for each target: HSC $i$- or $r$-band imaging \citep{aiharaHyperSuprimeCamSSP2018a}, HST WFC3/IR F110W imaging \citep{suessTextbackslashmathrmSQuIGGecLEStudyingQuenching2022,donofrioQuenchingTidalGas2025a,settonSQuIGGBuriedStar2025c}, or Dark Energy Camera Legacy Survey (DECaLS) $grz$ composites \citep{deyOverviewDESILegacy2019}. Each panel spans $6\arcsec\times6\arcsec$ with the orange cross marking the optical center. SDSS~J0224$-$0034, SDSS~J0227$-$0225, SDSS~J1235+0246, and SDSS~J1448+1010 are outlined in orange to identify the galaxies with clearly extended radio emission. Most detections are compact on few kpc scales, while these four sources exhibit extended structures.}
    \label{fig:detections}
\end{figure*}

\subsection{VLA Observations}
\label{label:vlaobs}

We observed 50 of the 52 targets with the VLA $C$-band receiver in the A-configuration during 2022 March under the project code 22A-362. $C$-band observations centered at 6\,GHz combine sensitivity to faint synchrotron emission with subarcsecond resolution. The observations used the standard $C$-band 3-bit continuum setup, providing continuous frequency coverage from 4.0 to 8.0\,GHz in two 2\,GHz-wide basebands centered at 5 and 7\,GHz, with dual polarization. Each baseband was divided into 128\,MHz spectral windows with 1\,MHz channelization. The flux, bandpass, and phase calibrators used for each object are listed in Appendix \ref{vlaobs} for clarity, with phase calibrators observed approximately every 6 minutes. We calibrated the data using the standard VLA calibration pipeline implemented in \texttt{CASA} \citep{mcmullinCASAArchitectureApplications2007,casateamCASACommonAstronomy2022a}. The remaining two galaxies, \PSBI\ and \PSBII, were previously observed in both the A- and B-configurations under projects 20A-188 and 20B-281 and presented in \citet{donofrioQuenchingTidalGas2025a}; further details of their observations and calibration are given in \citet{donofrioQuenchingTidalGas2025a}. 

Imaging was performed for all targets in \texttt{CASA} using Briggs weighting with \texttt{robust${}=0.5$} \citep{briggsHighFidelityDeconvolution1995}. We adopted a multiscale, multifrequency synthesis deconvolution with two Taylor terms to account for the frequency dependence of the continuum emission across the $C$-band bandwidth. To account for noncoplanar baseline effects across the wide fields of the images ($\approx$8$\arcmin$ on a side), particularly in fields containing bright sources away from the phase center, we applied the w-projection algorithm with 16 projection planes to all targets \citep{cornwellNoncoplanarBaselinesEffect2008}. The final continuum images reach a median local rms noise level of $4.9~\mu{\rm Jy\,beam^{-1}}$. The synthesized beams span $0\farcs29\times0\farcs28$ to $0\farcs58\times0\farcs54$, with a median beam size of $0\farcs35\times0\farcs30$. Across the redshift range of the sample, these resolutions correspond to physical scales of $1.8$--$4.2$\,kpc. The full table of final imaging properties is included in Appendix \ref{vlaobs}.

\begin{figure*}[ht!]
    \centering
    \includegraphics[width=\textwidth]
    {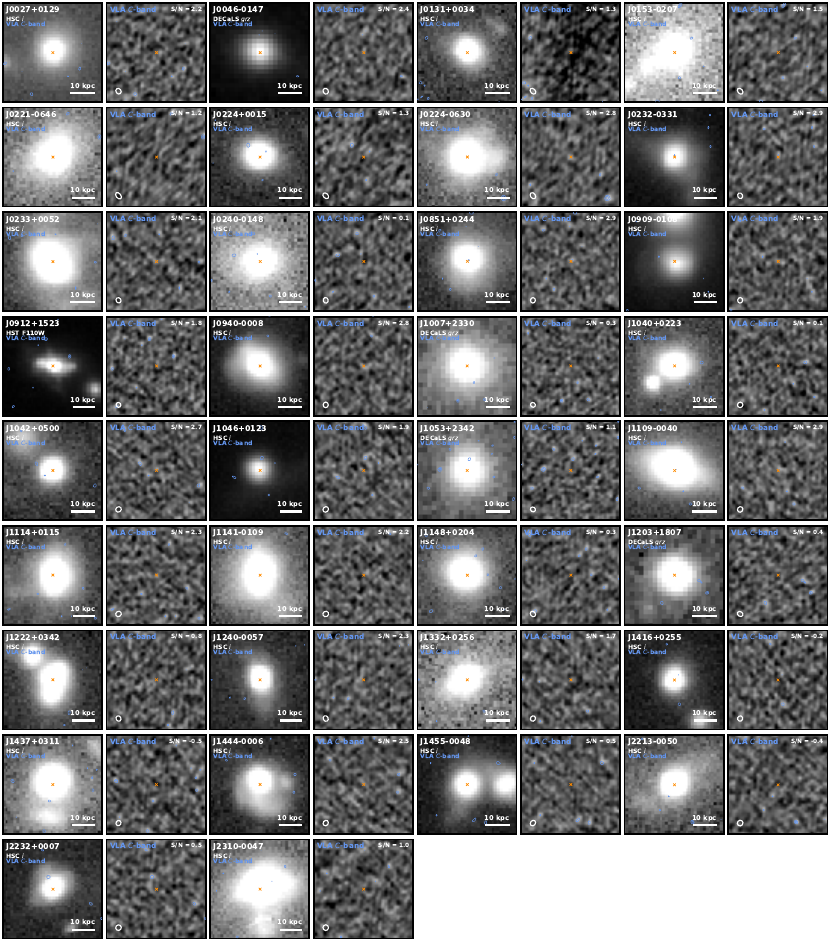}
    \caption{As in Figure~\ref{fig:detections}, but for the 34 undetected VLA $C$-band sources. Constraints on radio emission below the individual image sensitivity are obtained through the stacks presented in Figure~\ref{fig:stacks}.}
    \label{fig:nondetections}
\end{figure*}

%%%%%%%%%%%%%%%%%%%%%%%%%%%%%%%%%%%%%%%%%%%%%%%%%%%%%%%%%%%%%%%%%%%%%%%%%%%%%%%%%%%%%
%%%%%%%%%%%%%%%%%%%%%%%%%%%%%%%%% Data & Methods %%%%%%%%%%%%%%%%%%%%%%%%%%%%%%%%%%%%
%%%%%%%%%%%%%%%%%%%%%%%%%%%%%%%%%%%%%%%%%%%%%%%%%%%%%%%%%%%%%%%%%%%%%%%%%%%%%%%%%%%%%
\section{Analysis} \label{analysis}

\subsection{Radio Emission in \squiggle PSBs}
\label{label:radioem}

To assess how commonly radio activity occurs during the PSB phase, we examine the VLA 6\,GHz continuum emission. We detect VLA 6\,GHz continuum emission ($\geq3\sigma$) in 18 of the 52 \squiggle PSBs, corresponding to a detection fraction of $34.6^{+6.8}_{-6.3}\%$ (68\% binomial confidence interval). Figures~\ref{fig:detections} and \ref{fig:nondetections} show the radio imaging for the detected and undetected galaxies, respectively. We measure the source emission  using the \texttt{CASA IMFIT} task \citep{mcmullinCASAArchitectureApplications2007,casateamCASACommonAstronomy2022a}, fitting a single elliptical Gaussian to sources dominated by one component and multiple components where required by the observed structure, with integrated and peak flux densities reported in Table~\ref{tab:vla_props}.

The detected emission is generally centered on the optical galaxy and dominated by a single compact component near the resolution of the observations. Four sources (SDSS~J0224$-$0034, SDSS~J0227$-$0225, SDSS~J1235+0246, and SDSS~J1448+1010) exhibit clearly extended multicomponent structures; the latter was previously identified as having an extended structure \citep{donofrioQuenchingTidalGas2025a}. Resolved, extended morphologies provide evidence for AGN powered jets or dual AGN, whereas compact emission can arise from unresolved jets, recent star formation, or low luminosity accretion \citep[e.g.,][]{condonRadioEmissionNormal1992,murphyCalibratingExtinctionfreeStar2011,odeaCompactSteepSpectrum1998,heckmanCoEvolutionGalaxiesSupermassive2014,hardcastleCrostonRadioGalaxiesFeedback2020}. We examine the resolved morphology and origin of the radio emission further in Sections~\ref{label:morph} and \ref{label:origin}, respectively.

\begin{figure*}[ht!]
    \centering
    \includegraphics[width=0.925\textwidth]{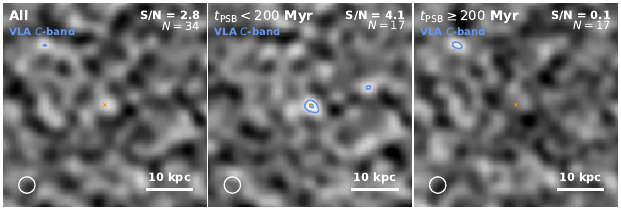}
    \caption{Mean-stacked VLA $C$-band images of all undetected sources and the subsets with $t_{\rm PSB}<200$\,Myr and $t_{\rm PSB}\geq200$\,Myr. VLA 6\,GHz contours are drawn at $3,4,5\sigma$. Significant stacked radio emission is detected only among the younger galaxies.}
    \label{fig:stacks}
\end{figure*}

Radio emission may also be present below the sensitivity of the individual images, therefore we stack the images of the 34 undetected sources. We register the undetected sources on their optical positions and convolve them to a common synthesized beam to form unweighted average stacks (Figure~\ref{fig:stacks}). Equal weighting is appropriate because it measures the mean emission of the galaxies themselves, rather than preferentially weighting  the few fields with the lowest rms. The full stack of undetected objects contains only a marginal central signal of $2.8\sigma$. Dividing the sample by PSB age, however, yields a $4.1\sigma$ detection for the 17 galaxies with $t_{\rm PSB}<200\,\mathrm{Myr}$, with a mean central flux density of $5.49\pm1.35\,\mu\mathrm{Jy\,beam^{-1}}$. The stack of the 17 older galaxies shows no significant emission ($0.2\pm1.5\,\mu\mathrm{Jy\,beam^{-1}}$). This result indicates that faint radio emission persists statistically below the survey threshold among the younger PSBs. We include these measurements when examining the dependence of radio activity on PSB age in Section~\ref{sec:age}.

Table~\ref{tab:vla_props} includes the individual radio luminosities of the sample (see Section~\ref{sec:lower} for further discussion). Converting the observed 6\,GHz flux densities to 1.4\,GHz assuming a spectral index of $\alpha=-0.7$ (appropriate for optically thin synchrotron emission from typical radio AGN, see Section~\ref{label:origin}), the detections span $\log(L_{1.4\,\mathrm{GHz}}/\mathrm{W\,Hz^{-1}})\approx23.0$--$25.0$. The detected luminosities therefore extend across roughly two orders of magnitude and reach more than an order of magnitude below those accessible to previous FIRST studies of \squiggle \citep{greeneRoleActiveGalactic2020} and other PSB samples \citep[e.g.,][]{luoMultiwavelengthEvaluationAGN2026,patilTracingRadioAGNDriven2026}, uncovering a substantial population of faint radio sources that was inaccessible to shallower surveys. 

\subsection{Radio Continuum Morphology}
\label{label:morph}

To determine whether the detected radio emission arises from compact nuclear sources or multicomponent structures, we characterize the 6\,GHz morphology of all 18 VLA detections. Along with the peak and integrated flux densities, Table~{\ref{tab:vla_props}} reports the deconvolved source properties. Comparable flux values indicate emission consistent with the synthesized beam, whereas an integrated flux substantially larger than the peak indicates emission distributed over multiple beams. For sources that can be reliably deconvolved from the beam using \texttt{IMFIT}, we convert the fitted major and minor axes to a circularized deconvolved diameter at the respective redshift. Six detections yield resolved diameters spanning $\approx0.5$--$10$\,kpc. For the remaining 12 sources, \texttt{IMFIT} reports that the radio emission is pointlike; we therefore report $2\sigma$ upper limits of $\lesssim0.1$--$1$\,kpc using the analytic point source approximation of \citet{condon1997}. These measurements characterize the extent of the fitted radio components and should not necessarily be interpreted as the total physical extent of especially morphologically complex sources.

Four galaxies (SDSS~J0224$-$0034, SDSS~J0227$-$0225, SDSS~J1235+0246, SDSS~J1448+1010) show particularly clear extended or multicomponent radio structures (Figure~\ref{fig:detections}). These sources constitute $\approx$20\% of the radio detections and exhibit extended, potentially jet-like structures, although their morphologies alone do not distinguish jets from star formation or AGN-driven wind shocks; multifrequency VLA observations will help determine whether these structures are morphologically consistent with radio jets. The remaining detections exhibit emission that is either pointlike or marginally resolved. However, at the $\sim0\farcs3$--$0\farcs6$ resolution of the observations (corresponding to physical scales of approximately $2$--$4$\,kpc at $z\sim0.7$) emission that is unresolved may reveal spatial structure at higher resolution. Circumnuclear star formation, compact AGN, and young or confined radio jets can all remain unresolved on these scales \citep[e.g.,][]{odeaCompactSteepSpectrum1998,heckmanCoEvolutionGalaxiesSupermassive2014,hardcastleCrostonRadioGalaxiesFeedback2020,patilTracingRadioAGNDriven2026}. The four extended systems therefore provide a lower limit on the incidence of multicomponent sources, while sources that appear pointlike at the current resolution may reveal resolved structure in higher-resolution observations.

\begin{figure*}[ht!]
    \centering
    \includegraphics[width=0.9\textwidth]{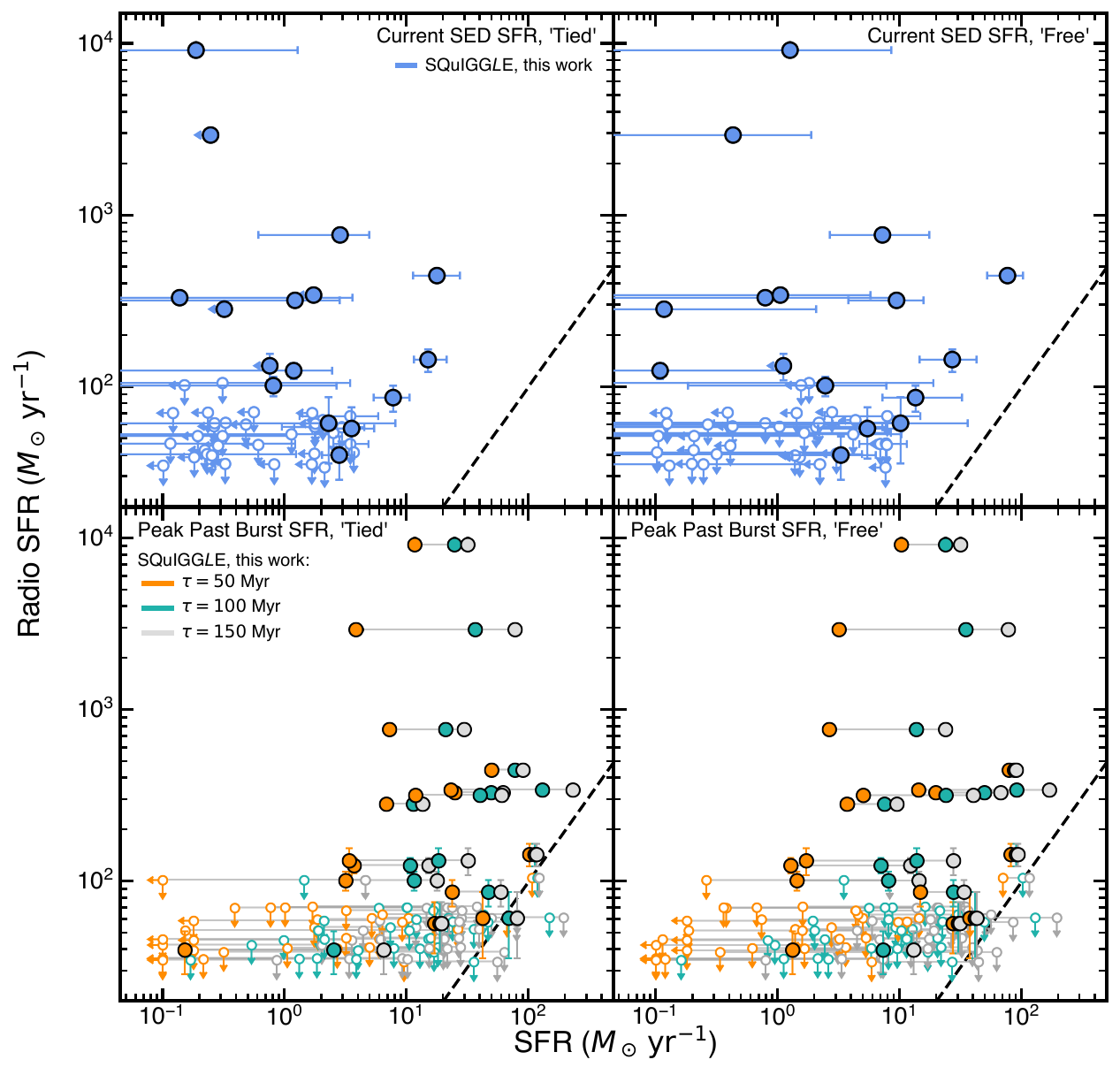}
    \caption{
    Comparison of radio-inferred SFRs with independent estimates of ongoing and recently declining star formation in \squiggle. The top row compares these values with the 10\,Myr-averaged SED SFRs \citep{settonSQuIGGBuriedStar2025c}, while the bottom row compares them with the radio emission predicted from the SFHs after allowing the starburst contribution to exponentially decline with radio decline timescales $\tau=50$, 100, and 150\,Myr. The left and right columns show the `Tied' and `Free' attenuation models of \citet{settonSQuIGGBuriedStar2025c}, respectively. The dashed line marks the one-to-one relation. The radio SFRs generally exceed both the current SED SFRs and the contribution predicted from the fading starburst.
    }
    \label{fig:sfrs}
\end{figure*}

\subsection{The Origin of the Radio Emission}
\label{label:origin}

Radio continuum emission associated with star formation is produced primarily by synchrotron emitting cosmic rays accelerated by core collapse supernovae, with a smaller thermal free--free contribution \citep[e.g.,][]{condonRadioEmissionNormal1992,murphyFarInfraredRadioCorrelationHigh2009,murphyCalibratingExtinctionfreeStar2011,tabatabaeiRadioSpectralEnergy2017}. We first test whether ongoing star formation can account for the observed radio emission by converting $L_{1.4\,{\rm GHz}}$ into a radio-inferred SFR (Equation~17 of \citealt{murphyCalibratingExtinctionfreeStar2011}), retaining $\alpha=-0.7$ as a reasonable approximation for synchrotron dominated star forming galaxies \citep[e.g.,][]{condonRadioEmissionNormal1992}, under the assumption that all of the radio emission is powered by star formation, and comparing it with the 10\,Myr-averaged SED SFRs described in Section~\ref{label:squiggleprops}. The top row of Figure~\ref{fig:sfrs} shows this comparison for the `Tied' and `Free' attenuation models of \citet{settonSQuIGGBuriedStar2025c} (see Section~\ref{label:squiggleprops}). For the VLA detections the SFR from radio emission exceeds the SED-derived value by more than a factor of three in every galaxy under both dust models, with median excess factors of $\approx$200 and $\approx$75 for the `Tied' and `Free' models, respectively. Even relative to the $3\sigma$ upper bounds, the sources generally retain a radio excess above the SED expectation from even the obscured SFR model. A $>$3$\times$ offset is larger than the scatter commonly allowed when identifying radio emission in excess of that powered by star formation and is comparable to the criterion commonly used to identify AGN-dominated radio sources \citep[e.g.,][]{barisicStellarDynamicsStarformation2017}. For three of the most CO-luminous galaxies, Pa$\alpha$ measurements likewise imply obscured SFRs below those required to reproduce their observed radio luminosities \citep{settonSpatiallyResolvedPaalpha2026}. The prevalence of radio excess across the sample indicates that ongoing star formation, including heavily obscured activity, is unlikely to dominate the observed radio emission.

The instantaneous SFR does not, however, account for synchrotron emission associated with the preceding starburst, which can persist after the rapid decline in star formation \citep[e.g.,][]{condonRadioEmissionNormal1992,murphyCalibratingExtinctionfreeStar2011,arango-toroProbingTimescale142023}. We therefore extend the prescription introduced in \citet{donofrioQuenchingTidalGas2025a} to the full sample using the updated SFHs from \citet{settonSQuIGGBuriedStar2025c}. For each galaxy, we estimate the radio emission associated with the SFR at the onset of quenching and allow this contribution to decline exponentially with characteristic timescales of $\tau=50$, 100, and 150\,Myr, which bracket plausible fading timescales for synchrotron emission from declining star formation \citep[e.g.,][]{condonRadioEmissionNormal1992,murphyCalibratingExtinctionfreeStar2011,arango-toroProbingTimescale142023}. These describe radio fading rather than the SFR decline itself as a prolonged burst and delayed supernova injection could broaden the radio response beyond this simplified prescription.

\begin{figure}[ht!]
    \centering
    \includegraphics[width=0.95\columnwidth]{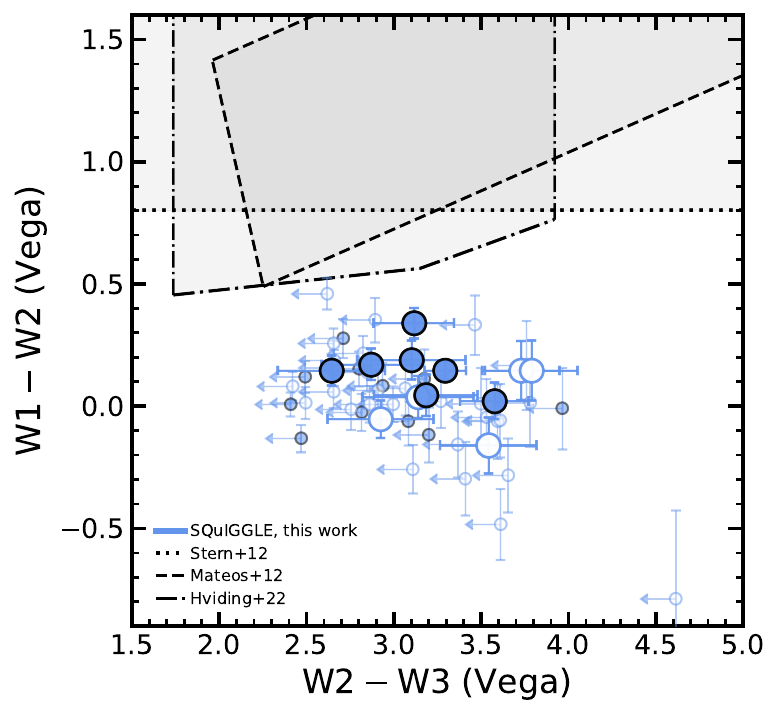}
    \caption{Observed-frame WISE $W1-W2$ versus $W2-W3$ colors for the \squiggle VLA sample. Filled blue circles denote VLA detections and open blue circles denote undetected VLA sources; larger symbols have $>3\sigma$ measurements in W1, W2, and W3, while arrows indicate color limits. The AGN selection boundaries of \citet{sternMidInfraredSelectionAGN2012}, \citet{mateosUsingWISEColoursIdentify2012}, and \citet{hvidingNewInfraredCriterion2022} are overlaid. None of the galaxies with securely measured three-band colors occupies the mid-infrared AGN selection regions, indicating that the integrated WISE emission is not dominated by a luminous dusty AGN.}
    \label{fig:wise}
\end{figure}

The bottom row of Figure~\ref{fig:sfrs} compares these predictions with the observed radio SFRs. The $\tau=150$\,Myr model provides the most permissive case, allowing substantial starburst-powered radio emission to persist well into the PSB phase. Nevertheless, most detections exceed the predictions from both models for even the longest decay time, with a median excess of approximately a factor of five in both cases for the longest timescale. Only the faintest radio measurements approach the range expected from the fading starburst. As noted in \citet{donofrioQuenchingTidalGas2025a} the exact normalization of this comparison remains sensitive to the adopted radio--SFR calibration and the simplified exponential fading prescription among other factors \citep[e.g.,][]{murphyFarInfraredRadioCorrelationHigh2009,tabatabaeiRadioSpectralEnergy2017,arango-toroProbingTimescale142023}. However, the persistence of the excess makes residual synchrotron emission from the recent burst unlikely to explain the majority of the detections.

\begin{figure}[t!]
    \centering
    \includegraphics[width=0.95\columnwidth]
    {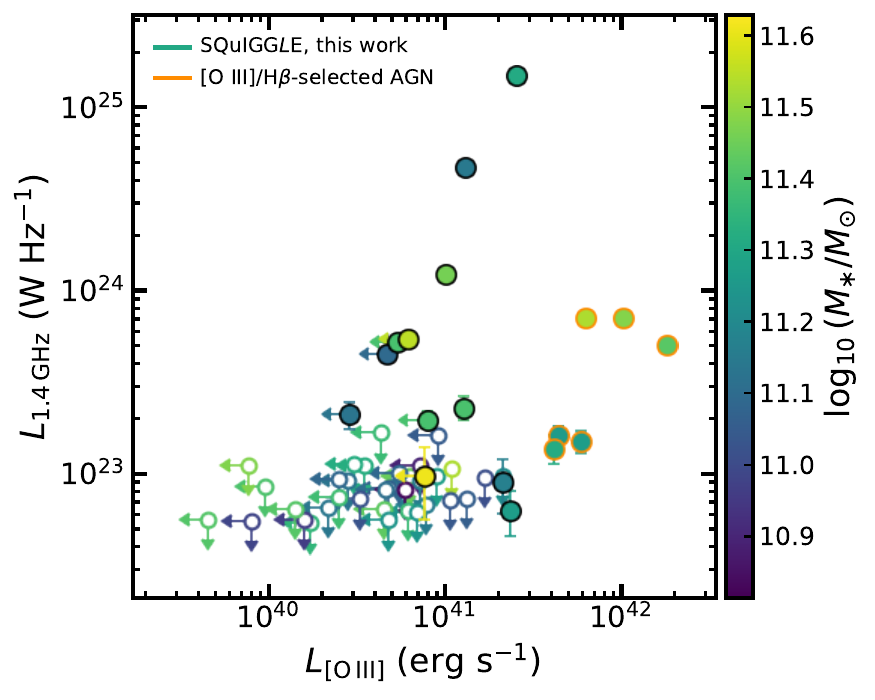}
    \caption{
    Radio luminosity versus \oiii luminosity for the
    \squiggle VLA sample. Orange outlines identify the six galaxies satisfying the \oiiihb AGN selection of \citet{greeneRoleActiveGalactic2020}, consistent with most of the radio population lacking strong optical AGN signatures.}
    \label{fig:OIII}
\end{figure}

The available ancillary diagnostics provide little evidence for a widespread population of luminous, radiatively efficient AGN. Figure \ref{fig:wise} shows the observed Wide-field Infrared Survey Explorer (WISE) colors of the VLA sample. All 52 galaxies are detected at ${\rm S/N}>3$ in W1 and W2, while only 12 have secure W3 measurements. None satisfies the W1--W2 AGN criterion of \citet{sternMidInfraredSelectionAGN2012}, and none of the galaxies with secure three-band photometry lies within the AGN regions defined by \citet{mateosUsingWISEColoursIdentify2012} or \citet{hvidingNewInfraredCriterion2022}. The integrated mid-infrared emission is therefore likely not dominated by a population of luminous, radiatively efficient AGN whose dusty tori dominate the host galaxy emission, consistent with the comparatively blue WISE colors reported for lower-redshift PSBs \citep[e.g.,][]{luoMultiwavelengthEvaluationAGN2026}. The SED modeling of \citet{settonSQuIGGBuriedStar2025c} nevertheless suggests a weak torus contribution in some sources (most apparent in SDSS~J1448+1010), consistent with nuclear infrared emission diluted by the host galaxy in the WISE colors. Similarly, Figure~\ref{fig:OIII} shows only six VLA targets satisfy the optical \oiiihb AGN selection of \citet{greeneRoleActiveGalactic2020}, and no clear trend between $L_{1.4\,\rm GHz}$ and $L_{\rm O\,III}$ exists. Among the nine gas-rich galaxies with CO(5--4) and CO(2--1) measurements, although most of this subset contains at least moderately excited molecular gas potentially reflecting heating associated with AGN activity, we find no apparent relation between radio luminosity and the excitation ratio \citep{donofrioCOExcitation2026}. We present this comparison in Appendix~\ref{radio_r52} because the subset is small and not representative of the full VLA sample. Merger driven shocks could also contribute to the enhanced radio incidence at higher stellar masses (see Section~\ref{sec:lower}), although the predominantly compact emission observed for those objects differs from the extended radio bridges modeled in collisional systems \citep[e.g.,][]{lisenfeldShockAccelerationRelativistic2010}.

\begin{figure*}[ht!]
    \centering
    \includegraphics[width=0.99\textwidth]
    {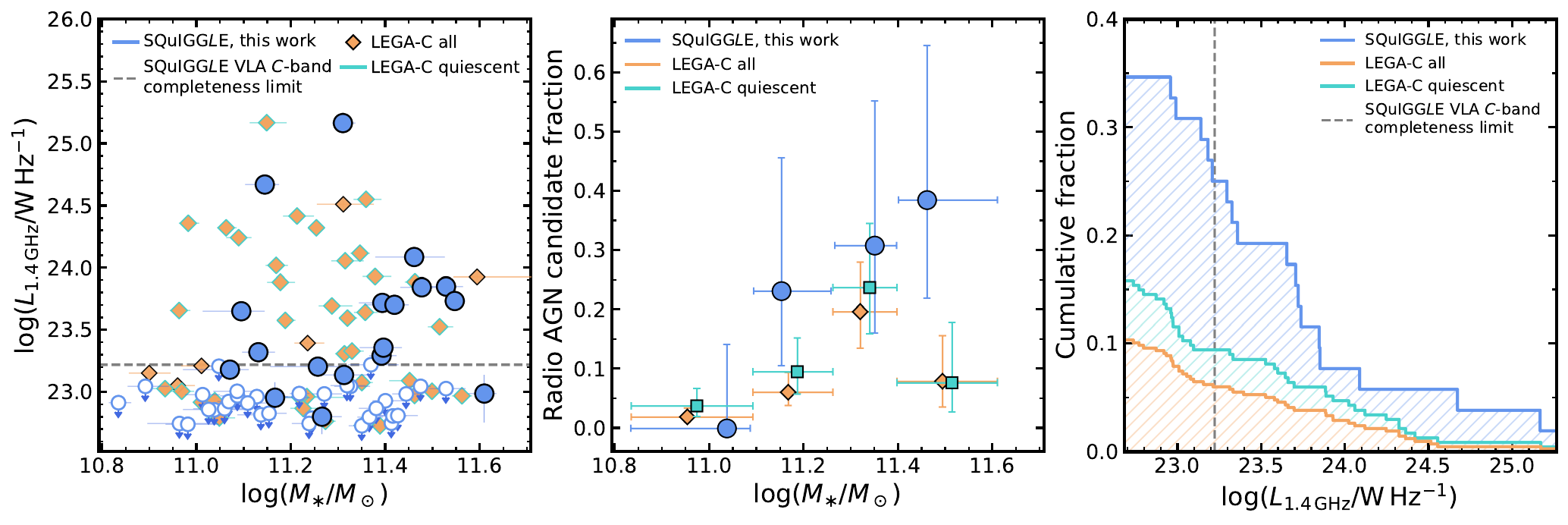}
    \caption{Comparison of radio AGN in \squiggle (blue) and the mass- and redshift-matched LEGA-C samples comprising solely quiescent radio AGN candidates (green) and the full population (star-forming and quiescent) radio AGN candidates (orange). Left: 1.4\,GHz radio luminosity versus stellar mass. The gray line marks the \squiggle VLA $C$-band completeness limit. Middle: radio AGN candidate fraction above the common completeness threshold of $1.7\times10^{23}\,\mathrm{W\,Hz^{-1}}$. Right: cumulative detection fraction of each population as a function of radio luminosity. Above the common threshold, radio AGN detections are approximately 4.0 and 2.5 times more frequent in \squiggle than in the full and quiescent LEGA-C samples, respectively, with the excess driven primarily by lower luminosity sources.}
    \label{fig:Lradio_Mstar}
\end{figure*}

Taken together, the general radio excess relative to both the current SFR and the fading contribution expected from the recent starburst favors an AGN contribution to most of the detected radio emission, while allowing star formation to contribute to some of the faintest measurements. The absence of corresponding strong mid-infrared or optical AGN signatures is not in tension with this interpretation as low-excitation, radiatively inefficient AGN can direct a large fraction of their energetic output into compact radio jets while remaining weak in conventional radiative diagnostics \citep[e.g.,][]{bestHostGalaxiesRadioloud2005,heckmanCoEvolutionGalaxiesSupermassive2014,hardcastleCrostonRadioGalaxiesFeedback2020}. Although the 12 brightest radio detections exceed the definition dependent radio-loud threshold ($L_{1.4\,{\rm GHz}}\gtrsim1.7\times10^{23}\,\mathrm{W\,Hz^{-1}}$) from \citet{bestHostGalaxiesRadioloud2005}, the remaining six extend below this limit (Table~\ref{tab:vla_props}). The combination of strong radio excesses and weak radiative AGN signatures is consistent with a population of comparatively low accretion rate, mechanically dominated AGN. Higher resolution multifrequency VLA observations can provide an additional test of this picture by resolving compact radio structures and spectral indices, helping distinguish compact cores and steep-spectrum jet emission from more diffuse star-forming disks \citep[e.g.,][]{odeaCompactSteepSpectrum1998,hardcastleCrostonRadioGalaxiesFeedback2020}.

%%%%%%%%%%%%%%%%%%%%%%%%%%%%%%%%%%%%%%%%%%%%%%%%%%%%%%%%%%%%%%%%%%%%%%%%%%%%%%%%%%%%%
%%%%%%%%%%%%%%%%%%%%%%%%%%%%%%%%% Results & Discussion %%%%%%%%%%%%%%%%%%%%%%%%%%%%%%
%%%%%%%%%%%%%%%%%%%%%%%%%%%%%%%%%%%%%%%%%%%%%%%%%%%%%%%%%%%%%%%%%%%%%%%%%%%%%%%%%%%%%
\section{Results and Discussion} \label{Results}

\subsection{Enhanced Lower Luminosity Radio Activity in \squiggle PSBs}
\label{sec:lower}

\begin{figure*}[ht!]
    \centering
    \includegraphics[width=0.99\textwidth]{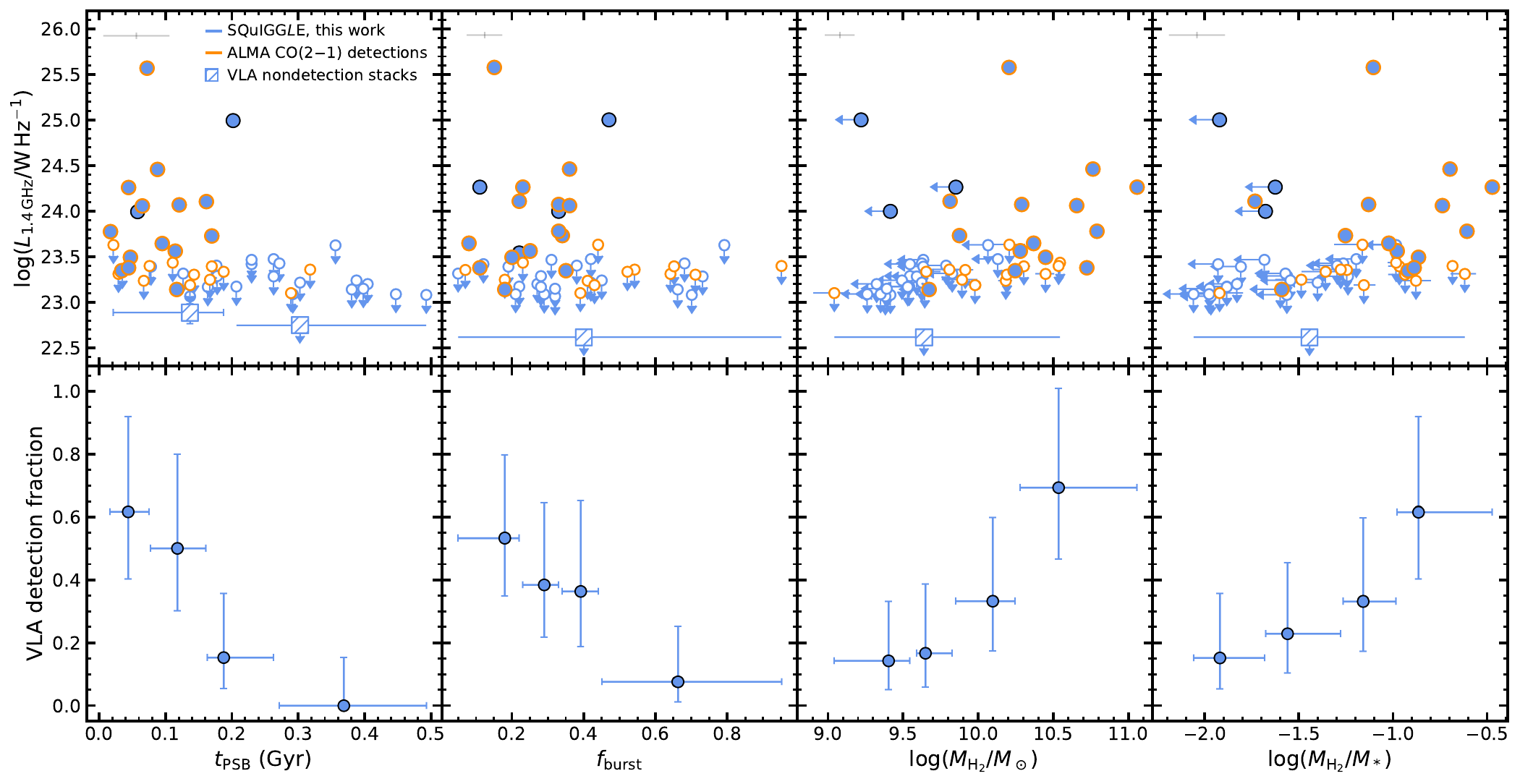}
    \caption{Radio activity against properties associated with the recent burst and subsequent molecular gas evolution. The upper row shows the 1.4\,GHz radio luminosities as functions of PSB age $t_{\rm PSB}$, recent burst mass fraction $f_{\rm burst}$, molecular gas mass $M_{\rm H_2}$, and molecular gas fraction $M_{\rm H_2}/M_*$. Orange outlines identify galaxies detected in ALMA CO(2--1) \citep{bezansonNowYouSee2022,settonSQuIGGBuriedStar2025c}. The gray crosses show representative uncertainties. Hatched squares show the stacked measurements of the undetected VLA sources; in the $t_{\rm PSB}$ panel, this is divided into younger and older populations (see Section~\ref{label:radioem}). The lower row presents the corresponding VLA detection fractions, with vertical bars showing 68\% Poisson confidence intervals. Radio detections occur predominantly among young, gas-rich galaxies.    
    }
    \label{fig:squigglecomp}
\end{figure*}

To determine whether radio activity is enhanced in PSBs, we compare the \squiggle measurements with a mass- and redshift-matched sample from the Large Early Galaxy Astrophysics Census (LEGA-C; \citealt{vanderwelVLTLEGACSpectroscopic2016}), using 3\,GHz radio measurements from \citet{smolcicVLACOSMOS3GHz2017} (converted to 1.4\,GHz assuming $\alpha=-0.7$ to match our measurements) and stellar masses from \citet{nersesianMoreBetterStrong2025}. Figure \ref{fig:Lradio_Mstar} places the \squiggle VLA measurements in the context of coeval massive quiescent galaxies from LEGA-C. We restrict the LEGA-C parent sample to the redshift and stellar mass ranges spanned by the VLA targets ($0.60<z<0.80$ and $10.8<\log(M_*/M_\odot)<11.6$). The resulting comparison sample contains 417 LEGA-C galaxies, of which 234 are classified as quiescent from their rest-frame $UVJ$ colors; the full sample contains both quiescent and star-forming galaxies. Following \citet{barisicStellarDynamicsStarformation2017}, we identify candidate radio AGN by requiring the observed radio luminosity exceed by a factor of three that expected from their ultraviolet+infrared SFRs, allowing for scatter in the radio--SFR relation. This comparison places the \squiggle PSBs in the context of older galaxies further along in their quiescent evolution \citep{kaushalCensusStarFormation2024}.

For the uniformly complete comparison between the complete populations in the middle panel of Figure~\ref{fig:Lradio_Mstar}, we adopt the shallowest \squiggle luminosity limit, $L_{1.4\,{\rm GHz}} = 1.7\times10^{23}\,\mathrm{W\,Hz^{-1}}$. We divide the \squiggle sample into four stellar mass bins containing approximately equal numbers of targets and apply the same boundaries to the LEGA-C samples. The vertical uncertainties on the binned fractions represent 68\% Poisson confidence intervals. Above this threshold, 12/52 ($23.1^{+8.7}_{-6.5}\%$) \squiggle galaxies are detected as radio AGN, compared with 24/417 ($5.8^{+1.4}_{-1.2}\%$) for the full LEGA-C comparison sample and 21/234 ($9.0^{+2.4}_{-1.9}\%$) for the quiescent subset. The \squiggle detection fraction is therefore approximately four times higher than in the full LEGA-C population and 2.5 times higher than in the more directly comparable quiescent population; two-sided Fisher exact tests indicate that the \squiggle sample differs by $3.8\sigma$ and $2.7\sigma$ for the full and quiescent LEGA-C samples, respectively. The enhancement is most apparent toward the higher mass portion of the sample, although the small numbers in individual mass bins and the strong underlying dependence of radio AGN incidence on stellar mass limit a more detailed interpretation \citep[e.g.,][]{bestHostGalaxiesRadioloud2005,heckmanCoEvolutionGalaxiesSupermassive2014,sabaterLoTSSViewRadio2019}. The cumulative radio fraction remains higher in \squiggle across the luminosity range shown in Figure~\ref{fig:Lradio_Mstar}, although the small numbers of sources at the highest luminosities limit the significance of this difference. The enhanced radio incidence in \squiggle is particularly apparent at lower radio luminosities. This behavior is consistent with comparatively weak or intermittent radio episodes occurring more frequently during the PSB phase, although a single epoch census cannot distinguish an increased duty cycle from longer lived episodes or a systematically different radio luminosity distribution \citep[e.g.,][]{bestHostGalaxiesRadioloud2005,hardcastleCrostonRadioGalaxiesFeedback2020,patilTracingRadioAGNDriven2026}.

\begin{figure}[ht!]
    \centering
    \includegraphics[width=0.99\columnwidth]{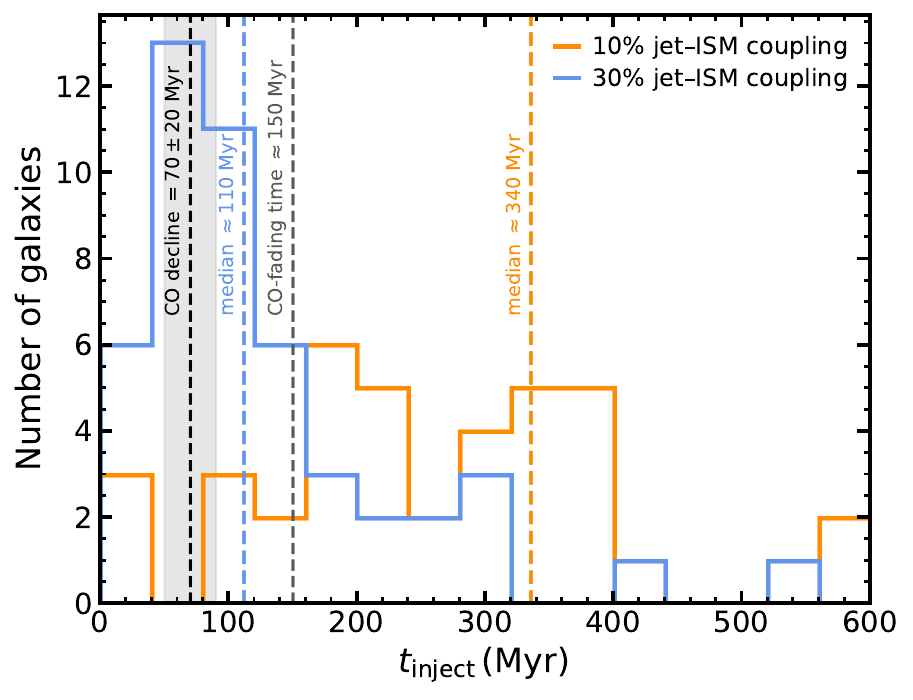}
    \caption{
    Estimated energy injection times for supplying an amount of mechanical energy comparable to the approximate molecular gas binding energy in the \squiggle galaxies. Orange and blue curves show the distributions for assumed coupling efficiencies of 10\% and 30\%, respectively. The dashed lines mark the corresponding median injection times. The black dotted line and gray band indicate the molecular gas decline timescale of $70\pm20$\,Myr measured for the \squiggle sample, while the gray dotted line marks approximately two e-folding times ($150$\,Myr), the timescale over which the molecular gas is expected to clear \citep{settonSQuIGGBuriedStar2025c}.
    }
    \label{fig:energy}
\end{figure}

At the full depth of the observations, radio emission is detected in 18/52 ($\approx$35\%) \squiggle galaxies. This is substantially higher than the $\approx$4\% radio AGN fraction previously measured with FIRST for the \squiggle sample \citep{greeneRoleActiveGalactic2020}. The difference is simply attributable to radio depth, where FIRST probes only the most luminous \squiggle sources ($L_{1.4\,{\rm GHz}}\sim(1$--$3.4)\times10^{24}\ {\rm W\,Hz^{-1}}$) whereas the presented VLA 6~GHz observations extend more than an order of magnitude lower in luminosity. Restricting the VLA sample to $L_{1.4\,{\rm GHz}}>10^{24}\ {\rm W\,Hz^{-1}}$ leaves only three detections which is broadly consistent with the \squiggle result from FIRST, the fewer than 8\% of PSBs at $0.48<z<1.2$ with radio counterparts \citep{verganiKGalaxiesZCOSMOS2010}, and the $5\pm2\%$ radio AGN candidate detection rate measured above the same threshold for massive PSBs at $0.5<z<3$ by \citet{patilTracingRadioAGNDriven2026}. The predominantly compact radio morphologies reported in the latter study indicate that extended structures are uncommon at the available resolution. However, the observed incidence of extended structure itself is perhaps sensitivity dependent, with some of the extended systems in Figure~\ref{fig:detections} containing a component at substantially lower significance than the brighter counterpart. This suggests that deeper imaging could reveal additional extended structures among sources currently classified as compact. 

At lower redshift, \citet{shinRadioEmissionActive2011} identified radio emission in only 12/513 ($\approx2\%$) local PSBs using the $\sim1$ mJy FIRST limit, of which 11 were attributed to AGN activity. Consistently, \citet{nielsenDeterminingAverageSFR2012} found that the stacked FIRST signal from local PSBs was dominated by the $\sim$4\% of fields containing individually radio-bright sources. \citet{luoMultiwavelengthEvaluationAGN2026} detected 8/73 ($\approx11\%$) PSBs in the radio but found no significant enhancement in AGN dominated radio emission relative to stellar mass- and redshift-matched star-forming and quiescent galaxies. The slightly higher incidence measured among intermediate-redshift PSBs in shallower radio surveys may indicate modest redshift evolution in bright radio activity. X-ray selected studies likewise find no general excess of luminous, radiatively efficient AGN among massive PSBs at cosmic noon \citep[e.g.,][]{almainiNoEvidenceExcessAGN2025}. These differing incidence rates emphasize the dependence of an AGN census on radio depth, in particular. The matched LEGA-C comparison more directly indicates that the PSB phase at intermediate redshift is associated specifically with an enhanced incidence of moderate radio activity. This result is consistent with recurrent or relatively weak radio episodes accompanying the early PSB phase and potentially contributing to maintaining quiescence, rather than a powerful radio AGN initiating the original shutdown of star formation.

\subsection{When Does Radio Activity Occur During the PSB Phase?}
\label{sec:age}

Figure~\ref{fig:squigglecomp} examines how radio activity varies across the PSB process by comparing to quantities that trace the recent burst and the subsequent evolution of the cold gas. Among the VLA detections, $L_{1.4\,{\rm GHz}}$ spans more than two orders of magnitude without a clear trend with any of the properties. %Although limited by the small number of detections, partial Pearson tests show no significant correlation between radio luminosity and either $t_{\rm PSB}$ or $f_{\rm burst}$ ($|r_p|\leq0.25$; $p\geq0.36$). Restricting the gas property tests to the 12 galaxies detected in both radio continuum and CO similarly yields no significant correlation with either $M_{\rm H_2}$ or $M_{\rm H_2}/M_*$ ($|r_p|\simeq0.23$; $p\simeq0.50$). 
More pronounced differences emerge in the incidence of radio emission. VLA detections predominantly occupy gas-rich galaxies, with 12/20 CO detected galaxies also detected in the radio continuum, compared with 5/31 undetected CO sources (the remaining radio detection, SDSS~J1235+0246, lacks CO coverage; see Section~\ref{data}). The detection fraction also increases with molecular gas mass and molecular gas fraction, possibly reflecting the availability of cold gas to fuel nuclear activity. Radio detections preferentially occur in galaxies with smaller recent burst mass fractions ($f_{\rm burst}$) and larger total stellar masses (see Figure~\ref{fig:Lradio_Mstar}). This could indicate that systems which converted a smaller fraction of their stellar mass into the recent burst retain more fuel for subsequent nuclear activity, although $f_{\rm burst}$ is a relative rather than absolute burst mass and its association with radio incidence may reflect covariance with stellar and gas content or with PSB age. In particular, the stellar mass dependence is qualitatively consistent with the strong increase of radio AGN incidence toward massive galaxies \citep[e.g.,][]{bestHostGalaxiesRadioloud2005,heckmanCoEvolutionGalaxiesSupermassive2014,sabaterLoTSSViewRadio2019,greeneRoleActiveGalactic2020}. However, the narrow mass range and intrinsic spread in $L_{1.4\,{\rm GHz}}$ prevent a meaningful luminosity--stellar mass slope from being measured.

However there is a clear sequence with PSB age, with 16/32 systems (SDSS~J0044$-$0054 and SDSS~J1235+0246 lack updated PSB ages from the \citealt{settonSQuIGGBuriedStar2025c} models and are excluded; see Section~\ref{data}) with $t_{\rm PSB}\lesssim0.2$\,Gyr detected in the radio, compared with none at older ages. The stacks of undetected sources provide a complementary view below the individual image sensitivity. While the full stack of undetected sources contains only a marginal $2.8\sigma$ signal, the stack of galaxies with $t_{\rm PSB}<200$\,Myr is detected at $4.1\sigma$ (Figure~\ref{fig:stacks}). The age dependence therefore persists below the nominal detection threshold and is not driven by a small number of radio-luminous young galaxies. Radio activity therefore occurs predominantly among young, gas-rich PSBs, although the correlated age and gas content prevent us from isolating their individual associations with radio incidence. The concentration of radio emission within the first 200\,Myr after the suppression of the starburst overlaps with the rapid evolution of the molecular reservoirs in \squiggle, for which the molecular gas fraction declines on an exponential timescale of $70\pm20$\,Myr and the CO-bright reservoir is inferred to fade over approximately two e-folding times ($\lesssim150$\,Myr) if the observed age trend traces an evolutionary sequence \citep{settonSQuIGGBuriedStar2025c}. One possibility is that the starburst and rapid suppression of star formation, retained molecular gas, and subsequent radio activity are different consequences of the same dissipative event, such that the radio source is largely `along for the ride' rather than responsible for shutting down star formation \citep[e.g.,][]{wildTimingAGNStarbursts2010,yesufStarburstQuiescenceTesting2014,lanzStarFormationSuppression2016,almainiNoEvidenceExcessAGN2025}. Alternatively, radio activity initiated near the end of the burst could help maintain low star formation efficiency through turbulence or dynamical disturbance \citep[e.g.,][]{crotonManyLivesActive2006,alataloSuppressionStarFormation2015a,lanzStarFormationSuppression2016,hardcastleCrostonRadioGalaxiesFeedback2020,luoMultiwavelengthEvaluationAGN2026}.      

\subsection{Can Radio AGN Unbind the Remaining Molecular Gas?}
\label{sec:maintain}

We next ask whether the observed radio sources could inject sufficient mechanical energy to dynamically affect the remaining molecular gas reservoirs of the \squiggle PSBs. Figure~\ref{fig:energy} presents an order-of-magnitude energetic comparison intended to assess the plausibility of this scenario, rather than model the detailed interaction between radio jets and the multiphase ISM. Similar energy arguments have been used to examine whether AGN-driven shocks and turbulence can suppress star formation without necessarily removing the molecular gas \citep[e.g.,][]{alataloSuppressionStarFormation2015a,lanzStarFormationSuppression2016,smercinaFallResolvingMolecular2022b}. We characterize the energy required to dynamically perturb and unbind the molecular reservoirs using $E_{\rm required} \sim M_{\rm H_2}\sigma_{*}^{2}$ where $M_{\rm H_2}$ is the molecular gas mass from the ALMA CO(2--1) observations \citep{bezansonNowYouSee2022,settonSQuIGGBuriedStar2025c} and $\sigma_{*}$ is the stellar velocity dispersion \citep{suessTextbackslashmathrmSQuIGGecLEStudyingQuenching2022}. We adopt $\sigma_{*}$ as a common characteristic velocity scale as reliable gas velocity dispersions are unavailable for the undetected CO(2--1) sources; for the CO detections the gas and stellar dispersions are comparable (see \citealt{settonSQuIGGBuriedStar2025c} for the CO line profiles of the sample). This approximation follows the usual kinetic energy scaling, with geometric and dimensional factors absorbed into the order unity uncertainty. The resulting $E_{\rm required}$ should be interpreted as a characteristic energy scale for dynamically disturbing the gas rather than a direct measurement of its turbulent or binding energy. We define the corresponding energy injection timescale as $t_{\rm inject} \sim E_{\rm required}\big/
\left(\epsilon P_{\rm jet}\right)$, where $\epsilon$ is the jet--ISM coupling efficiency and the mechanical jet power $P_{\rm jet}$ is estimated from the 1.4~GHz luminosity following the radio--jet power calibration in \citet{heckmanCoEvolutionGalaxiesSupermassive2014}. We consider coupling efficiencies of 10\% and 30\%, broadly spanning moderate to efficient coupling between compact jets and a dense, clumpy ISM \citep[e.g.,][]{mukherjeeRelativisticJetFeedback2016,mukherjeeRelativisticJetFeedback2018,hardcastleCrostonRadioGalaxiesFeedback2020}. For clarity we show timescales derived from the $M_{\rm H_2}$ and $P_{\rm jet}$ upper limits alongside the measurements to trace this approximate timescale across the full evolutionary sequence, particularly because the older PSBs are preferentially undetected in both CO and the radio. We compare these estimates with the $70\pm20$~Myr exponential decline in molecular gas fraction and the $\lesssim150$~Myr timescale over which the remaining CO reservoir is expected to fade if the observed age trend represents an evolutionary sequence \citep{settonSQuIGGBuriedStar2025c}. 

Figure~\ref{fig:energy} shows that the median injection times are $\approx$340 and $\approx$110\,Myr for coupling efficiencies of 10\% and 30\%, respectively. At 30\% coupling the characteristic injection time is comparable to the observed evolution of the molecular reservoirs, with $\approx$70\% of PSBs reaching the adopted energy scale within 150\,Myr, compared with just $\approx$15\% at 10\% coupling. The higher efficiency approximation is therefore comparable in order of magnitude to both the $70\pm20$~Myr exponential decline in CO luminosity and the $\lesssim150$~Myr interval over which the population becomes predominantly CO-faint \citep{settonSQuIGGBuriedStar2025c}. Under these assumptions, if the radio jets couple efficiently to the surrounding gas, they could supply the approximate energy required to unbind the molecular gas while its reservoirs are fading. Such behavior is characteristic of maintenance-mode feedback, in which mechanically dominated, comparatively low accretion rate AGN lower the central density of the remaining gas rather than ejecting the entire ISM in a single radiatively efficient quasar episode \citep[e.g.,][]{crotonManyLivesActive2006,fabianObservationalEvidenceAGN2012,heckmanCoEvolutionGalaxiesSupermassive2014,hardcastleCrostonRadioGalaxiesFeedback2020}. The compact radio morphologies found for most detections could facilitate particularly effective coupling within the central ISM, while the four extended sources if confirmed as jets can already demonstrate their presence during this evolutionary stage. In this interpretation, radio activity is not required to have caused the initial shutdown of the burst, but may instead preserve the suppressed state while the residual gas is displaced.

However, this interpretation remains conditional on simplifying assumptions. The calculation treats the observed radio luminosity as a tracer of constant jet power, whereas radio AGN are variable and episodic. For an intermittent source, $t_{\rm inject}$ represents the cumulative active time, and the elapsed time required to deliver the same energy increases according to the duty cycle \citep[e.g.,][]{hickoxBlackHoleVariability2014,heckmanCoEvolutionGalaxiesSupermassive2014,hardcastleCrostonRadioGalaxiesFeedback2020}. Moreover, AGN activity can lag the peak of the starburst \citep[e.g.,][]{wildTimingAGNStarbursts2010,yesufStarburstQuiescenceTesting2014,frenchAGNLightEchoes2023}. The radio activity observed during the PSB phase could therefore represent delayed feedback that contributed to the final shutdown of star formation, or a later maintenance episode that began only after quenching, possibilities that cannot be distinguished by present timing constraints. Neutral gas observations of the youngest, gas-rich \squiggle galaxies reveal $\sim$$200$--$400\,{\rm km\,s^{-1}}$ outflows with mass loss rates of order $40\,M_\odot\,{\rm yr^{-1}}$, whose velocities and energetics may retain a memory of the preceding starburst or an earlier AGN episode \citep{luoUbiquitousCoolOutflows2026}. The radio emission could possibly trace fading shocks from an earlier radiatively efficient AGN episode rather than ongoing jet activity, a possibility that the present single band observations cannot distinguish.

An alternative possibility is that some gas-rich objects are temporarily stalled rather than permanently quenched. JWST Pa$\alpha$ measurements of the three most CO-luminous \squiggle galaxies find dust-corrected SFRs of  $\sim30$--$70\,M_\odot\,{\rm yr^{-1}}$, although these may be overestimated by $\sim$30--35\% as shocks or AGN can contribute to the recombination-line emission \citep{settonSpatiallyResolvedPaalpha2026}. The gas-rich galaxies are therefore not completely passive, but their current SFRs remain $\sim$0.5--1.3\,dex below the reconstructed peaks of their bursts, consistent with a rapid decline that has not yet reached complete quiescence. More generally, a retained molecular reservoir may continue forming stars inefficiently or instead fuel a later rejuvenation episode \citep[e.g.,][]{settonSQuIGGBuriedStar2025c,cenciNaturePostStarburstGalaxies2025b}. The ubiquitous tidal structure in the gas-rich \squiggle systems further indicates recent mergers \citep{spilkerStarFormationSuppression2022,verricoMergerSignaturesAre2023,donofrioQuenchingTidalGas2025a}, while low-redshift post-mergers show that rapid quenching and renewed nuclear activity can occur over several hundred Myr following coalescence \citep{ellisonGalaxyEvolutionPostmerger2024a,ellisonGalaxyEvolutionPostmerger2025a}. Simulations predict PSB selections containing both genuinely quenched systems and galaxies undergoing temporary lulls or intermittent star formation \citep{cenciNaturePostStarburstGalaxies2025b}. \citet{savitchOriginsMolecularReservoirs2026} further suggest that recurrent jets can depress central gas densities and stall star formation while leaving fuel available for later activity. The current data therefore cannot distinguish conclusively between permanent maintenance of quiescence and a temporary interruption of star formation. They do show, however, that the observed radio sources have sufficient mechanical power to influence the remaining gas on the relevant evolutionary timescale, making jet-driven maintenance a physically viable component of the transition from the youngest gas-rich PSBs toward their gas-poor descendants.

%%%%%%%%%%%%%%%%%%%%%%%%%%%%%%%%%%%%%%%%%%%%%%%%%%%%%%%%%%%%%%%%%%%%%%%%%%%%%%%%%%%%%
%%%%%%%%%%%%%%%%%%%%%%%%%%%%%%%%% Conclusions %%%%%%%%%%%%%%%%%%%%%%%%%%%%%%%%%%%%%%%
%%%%%%%%%%%%%%%%%%%%%%%%%%%%%%%%%%%%%%%%%%%%%%%%%%%%%%%%%%%%%%%%%%%%%%%%%%%%%%%%%%%%%
\section{Conclusions}
\label{Conclusions}

We have presented deep, subarcsecond VLA $C$-band observations of 52 massive \squiggle PSBs at $z\sim0.7$. We detect 6\,GHz emission in 18/52 galaxies ($\approx$35\%), with luminosities spanning $\log(L_{1.4\,{\rm GHz}}/{\rm W\,Hz^{-1}})\simeq23.0$--$25.0$. Four sources show extended radio structures possibly consistent with compact and/or young radio jets, while the remaining detections are pointlike or marginally resolved at the $\sim$2--4\,kpc physical resolution of the observations. The measured radio luminosities generally exceed those expected from both the current SED-inferred SFRs and residual synchrotron emission from the preceding starburst by a factor of a few, even when allowing for heavily obscured star formation. Conversely, the WISE color diagnostics and \oiiihb measurements provide little evidence for a widespread population of luminous, radiatively efficient AGN. Together, these results favor an AGN dominated contribution to most of the detected radio emission, consistent with a population of comparatively radiatively weak, mechanically dominated sources, while allowing star formation to contribute to some of the faintest detections. 

Radio activity changes quickly across the post-burst age sequence, where each radio detection is found in a PSB with $t_{\rm PSB}<0.2$\,Gyr. The stacks of undetected sources extend this trend below the individual source sensitivity with the youngest galaxies yielding a detected signal, suggesting that the measured detection fraction remains sensitivity limited and still fainter radio activity may be present during the earliest PSB phase. Radio detections are also predominantly associated with galaxies that retain substantial molecular gas reservoirs \citep{bezansonNowYouSee2022,settonSQuIGGBuriedStar2025c}.

At the full depth of the VLA observations, the \squiggle detection fraction is considerably higher than the radio source fractions reported for other PSB samples. Much of this difference reflects the greater radio depth rather than an excess of the most powerful sources. When restricted to $L_{1.4\,{\rm GHz}}>10^{24}\,\mathrm{W\,Hz^{-1}}$, the \squiggle fraction becomes broadly consistent with previous shallower PSB surveys \citep[e.g.,][]{shinRadioEmissionActive2011,greeneRoleActiveGalactic2020,luoMultiwavelengthEvaluationAGN2026,patilTracingRadioAGNDriven2026}. Above a common completeness threshold $L_{1.4\,{\rm GHz}}=1.7\times10^{23}\,\mathrm{W\,Hz^{-1}}$, the fraction of \squiggle galaxies with radio emission is approximately four times the radio AGN candidate fraction in the full mass- and redshift-matched LEGA-C galaxy sample and approximately 2.5 times that in its quiescent subset. The post-burst enhancement is driven primarily by weaker radio activity rather than an overabundance of high power radio AGN. Deep radio observations therefore illustrate that weaker or intermittent radio episodes may be common during the early PSB phase, although a single epoch census cannot distinguish an enhanced duty cycle from differences in episode duration or luminosity. 

The radio sources are energetically capable of influencing the surviving molecular gas under sufficiently efficient jet--ISM coupling. For a coupling efficiency of 30\%, the injection timescale is comparable in order of magnitude to both the $70\pm20$\,Myr decline in CO luminosity and the $\lesssim150$\,Myr interval over which the galaxies are expected to clear their gas reserves \citep{settonSQuIGGBuriedStar2025c}. Multifrequency radio constraints on jet ages and powers, together with spatially resolved molecular gas kinematics, will provide more direct measurements of the relevant duty cycles and coupling efficiencies. Conversely, the radio activity may arise from the same gas-rich event that produced the starburst rather than initiating its rapid suppression. Nevertheless, sufficiently well coupled jets could subsequently help maintain inefficient star formation.

Together, these results identify faint radio activity as a common feature of the youngest, gas-rich \squiggle PSBs and show that its incidence declines over approximately the same evolutionary interval in which their CO-luminous reservoirs fade. Deeper radio observations of larger PSB samples will determine whether this activity becomes nearly ubiquitous at early PSB times or remains confined to a distinct subset. Multifrequency radio observations will help constrain spectral curvature and synchrotron ages, allowing the timing and duty cycles of individual jet episodes to be compared directly with the reconstructed SFHs \citep[e.g.,][]{myersSpanglerSynchrotronAging1985,harwoodSpectralAgeingLobes2013}. Spatially resolved comparisons with molecular gas kinematics can test whether the jets are dynamically coupled to the cold ISM, while dense gas tracers such as HCN, HCO$^+$, and HNC can constrain both the fraction of gas in the dense, star forming phase and its ionization state \citep[e.g.,][]{privonExcitationMechanismsHCNHCO2015,frenchStateMolecularGas2023c}. Connecting the timing and energetics of faint radio activity to the changing physical state of the molecular gas will determine whether these sources merely accompany rapid quenching or help sustain the transition toward quiescence.     
 
\section*{Acknowledgements}
V.R.D., J.S.S., and K.A.S. gratefully acknowledge support from the National Science Foundation under NSF-AAG Nos. 2407954 and 2407955. V.R.D. also acknowledges support provided by the NSF through award SOSPA 11-006 from the NRAO. K.L.S. gratefully acknowledges support from the Mitchell-Heep-Munnerlyn career enhancement chair at Texas A\&M University.

This paper makes use of data from the Karl G. Jansky Very Large Array obtained under programs 20A-188, 20B-281, and 22A-362. The National Radio Astronomy Observatory is a facility of the National Science Foundation operated under cooperative agreement by Associated Universities, Inc.

This research is based on observations made with the NASA/ESA Hubble Space Telescope and obtained from the Mikulski Archive for Space Telescopes at the Space Telescope Science Institute, which is operated by the Association of Universities for Research in Astronomy, Inc., under NASA contract NAS 5--26555. These observations are associated with programs 15436, 16201, and 16248. The specific observations analyzed can be accessed via \dataset[10.17909/9xpd-3752]{https://doi.org/10.17909/9xpd-3752} and \dataset[10.17909/p5bd-nn41]{https://doi.org/10.17909/p5bd-nn41}. Support for MAST for these data is provided by the NASA Office of Space Science via grant NAG5--7584 and by other grants and contracts.

The Hyper Suprime-Cam (HSC) collaboration includes the astronomical communities of Japan and Taiwan, and Princeton University. The HSC instrumentation and software were developed by the National Astronomical Observatory of Japan (NAOJ), the Kavli Institute for the Physics and Mathematics of the Universe (Kavli IPMU), the University of Tokyo, the High Energy Accelerator Research Organization (KEK), the Academia Sinica Institute for Astronomy and Astrophysics in Taiwan (ASIAA), and Princeton University. Funding was contributed by the FIRST program from the Japanese Cabinet Office, the Ministry of Education, Culture, Sports, Science and Technology (MEXT), the Japan Society for the Promotion of Science (JSPS), the Japan Science and Technology Agency (JST), the Toray Science Foundation, NAOJ, Kavli IPMU, KEK, ASIAA, and Princeton University.

This work is based in part on data collected at the Subaru Telescope and retrieved from the HSC data archive system, which is operated by the Subaru Telescope and Astronomy Data Center at the National Astronomical Observatory of Japan.

This research has made use of NASA's Astrophysics Data System.

\facility{HST (WFC3), SDSS, Subaru (HSC), VLA}

\software{
\texttt{astropy} \citep{astropycollaborationAstropyCommunityPython2013a,astropycollaborationAstropyProjectBuilding2018,astropycollaborationAstropyProjectSustaining2022a},
CASA \citep{casateamCASACommonAstronomy2022a},
\texttt{matplotlib} \citep{hunterMatplotlib2DGraphics2007}
}

\clearpage
\appendix
\twocolumngrid

\section{VLA Observational Properties} \label{vlaobs}

Here, we present Table~\ref{tab:cals}, which lists the calibrators and final image properties for the VLA observations of each source.

\begin{table*}[p]
\centering
\caption{VLA Calibrators and Image Properties.}
\label{tab:cals}
\begingroup
\scriptsize
%\fontsize{8.7pt}{8.3pt}\selectfont
\setlength{\tabcolsep}{-5pt}
\renewcommand{\arraystretch}{0.99}
\begin{tabular*}{0.9\textwidth}{@{\extracolsep{\fill}}lcccc@{}}
\hline
\hline
Object & Flux/bandpass calibrator & Phase calibrator &
Synthesized beam & rms \\
 & & & ($\arcsec\times\arcsec$) & ($\mu$Jy beam$^{-1}$) \\
\hline
SDSS~J0027+0129 & 3C48 & J0040-0146 & 0.36 $\times$ 0.31 & 5.1 \\
SDSS~J0044-0054 & 3C48 & J0040-0146 & 0.37 $\times$ 0.31 & 5.1 \\
SDSS~J0046-0147 & 3C48 & J0040-0146 & 0.37 $\times$ 0.32 & 5.1 \\
SDSS~J0131+0034 & 3C48 & J0125-0005 & 0.41 $\times$ 0.28 & 7.5 \\
SDSS~J0153-0207 & 3C48 & J0215-0222 & 0.41 $\times$ 0.29 & 4.7 \\
SDSS~J0221-0646 & 3C48 & J0215-0222 & 0.45 $\times$ 0.28 & 4.7 \\
SDSS~J0224+0015 & 3C48 & J0215-0222 & 0.41 $\times$ 0.29 & 4.8 \\
SDSS~J0224-0034 & 3C48 & J0215-0222 & 0.40 $\times$ 0.29 & 4.7 \\
SDSS~J0224-0630 & 3C48 & J0215-0222 & 0.41 $\times$ 0.30 & 4.7 \\
SDSS~J0227-0225 & 3C48 & J0239-0234 & 0.36 $\times$ 0.29 & 4.9 \\
SDSS~J0232-0331 & 3C48 & J0239-0234 & 0.36 $\times$ 0.29 & 4.8 \\
SDSS~J0233+0052 & 3C48 & J0239-0234 & 0.34 $\times$ 0.30 & 4.9 \\
SDSS~J0237+0123 & 3C48 & J0239-0234 & 0.34 $\times$ 0.30 & 5.0 \\
SDSS~J0240-0148 & 3C48 & J0239-0234 & 0.35 $\times$ 0.30 & 5.1 \\
SDSS~J0753+2403 & 3C286 & J0748+2400 & 0.29 $\times$ 0.28 & 4.7 \\
SDSS~J0851+0244 & 3C286 & J0909+0121 & 0.33 $\times$ 0.30 & 4.9 \\
SDSS~J0907+0423 & 3C286 & J0909+0121 & 0.32 $\times$ 0.30 & 4.7 \\
SDSS~J0909-0108 & 3C286 & J0909+0121 & 0.34 $\times$ 0.30 & 4.8 \\
SDSS~J0910+0218 & 3C286 & J0909+0121 & 0.34 $\times$ 0.31 & 4.6 \\
SDSS~J0912+1523 & 3C286 & J0854+2006 & 0.31 $\times$ 0.28 & 4.8 \\
SDSS~J0940-0008 & 3C286 & J0925+0019 & 0.34 $\times$ 0.31 & 4.8 \\
SDSS~J1007+2330 & 3C286 & J0956+2515 & 0.31 $\times$ 0.28 & 4.6 \\
SDSS~J1017-0003 & 3C286 & J1024-0052 & 0.34 $\times$ 0.31 & 4.7 \\
SDSS~J1040+0223 & 3C286 & J1058+0133 & 0.34 $\times$ 0.33 & 6.8 \\
SDSS~J1042+0500 & 3C286 & J1058+0133 & 0.36 $\times$ 0.31 & 7.1 \\
SDSS~J1046+0123 & 3C286 & J1058+0133 & 0.36 $\times$ 0.31 & 7.5 \\
SDSS~J1053+2342 & 3C286 & J1051+2119 & 0.32 $\times$ 0.28 & 4.7 \\
SDSS~J1109-0040 & 3C286 & J1058+0133 & 0.35 $\times$ 0.32 & 8.1 \\
SDSS~J1114+0115 & 3C286 & J1058+0133 & 0.36 $\times$ 0.32 & 8.1 \\
SDSS~J1141-0109 & 3C286 & J1150-0023 & 0.35 $\times$ 0.33 & 7.5 \\
SDSS~J1142+0006 & 3C286 & J1150-0023 & 0.36 $\times$ 0.32 & 7.3 \\
SDSS~J1148+0204 & 3C286 & J1150-0023 & 0.36 $\times$ 0.30 & 4.6 \\
SDSS~J1157+0132 & 3C286 & J1150-0023 & 0.36 $\times$ 0.30 & 4.7 \\
SDSS~J1203+1807 & 3C286 & J1158+2450 & 0.37 $\times$ 0.29 & 4.7 \\
SDSS~J1211+0240 & 3C286 & J1224+0330 & 0.35 $\times$ 0.30 & 4.6 \\
SDSS~J1222+0342 & 3C286 & J1224+0330 & 0.34 $\times$ 0.29 & 4.5 \\
SDSS~J1235+0246 & 3C286 & J1224+0330 & 0.35 $\times$ 0.30 & 4.6 \\
SDSS~J1240-0057 & 3C286 & J1224+0330 & 0.36 $\times$ 0.30 & 4.7 \\
SDSS~J1244+0248 & 3C286 & J1224+0330 & 0.34 $\times$ 0.30 & 4.7 \\
SDSS~J1332+0256 & 3C286 & J1354-0206 & 0.34 $\times$ 0.30 & 5.3 \\
SDSS~J1416+0255 & 3C286 & J1354-0206 & 0.33 $\times$ 0.31 & 5.4 \\
SDSS~J1436+0447 & 3C286 & J1505+0326 & 0.32 $\times$ 0.30 & 5.1 \\
SDSS~J1437+0311 & 3C286 & J1505+0326 & 0.33 $\times$ 0.30 & 5.5 \\
SDSS~J1444-0006 & 3C286 & J1505+0326 & 0.35 $\times$ 0.30 & 5.2 \\
SDSS~J1448+1010 & 3C286 & J1445+0958 & 0.46 $\times$ 0.37 & 2.0 \\
SDSS~J1449+0206 & 3C286 & J1505+0326 & 0.35 $\times$ 0.30 & 5.3 \\
SDSS~J1455-0048 & 3C286 & J1505+0326 & 0.37 $\times$ 0.30 & 5.4 \\
SDSS~J2202-0033 & 3C48 & J2218-0335 & 0.36 $\times$ 0.34 & 5.4 \\
SDSS~J2213-0050 & 3C48 & J2218-0335 & 0.36 $\times$ 0.34 & 5.3 \\
SDSS~J2232+0007 & 3C48 & J2218-0335 & 0.35 $\times$ 0.33 & 5.4 \\
SDSS~J2258+2313 & 3C48 & J2254+2445 & 0.58 $\times$ 0.54 & 2.0 \\
SDSS~J2310-0047 & 3C48 & J2323-0317 & 0.35 $\times$ 0.33 & 5.3 \\
\hline
\end{tabular*}
\begin{minipage}[t]{0.89\textwidth}
\vspace*{4pt}
\raggedright
{\fontsize{10pt}{10pt}\selectfont N\textsc{ote}}---The calibrators were used in the survey program 22A-362 and pilot programs 20A-188 and 20B-281. J1448+1010 and J2258+2313 use the same calibrator sets in their pilot A- and B-configuration observations.
\end{minipage}
\endgroup
\end{table*}

\section{Radio Luminosity and CO Excitation} \label{radio_r52}

\begin{figure}[h!]
    \centering
    \includegraphics[width=0.95\columnwidth]
    {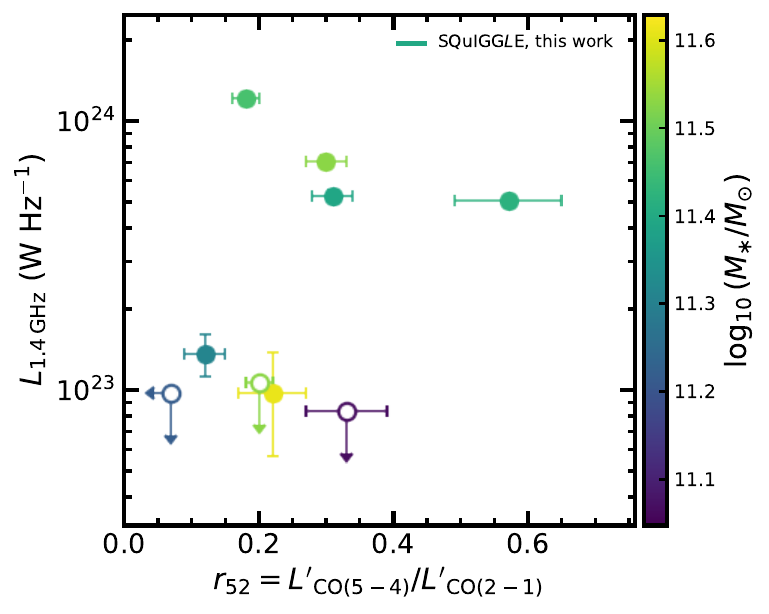}
    \caption{Radio luminosity versus the CO excitation ratio for the nine gas-rich
    \squiggle galaxies with both CO(5--4) and CO(2--1) coverage. We find no clear relationship between radio luminosity and molecular gas excitation in this subset of the presented VLA sample.
    }
    \label{fig:r52}
\end{figure}

Figure~\ref{fig:r52} compares the 1.4\,GHz radio luminosity with the CO excitation ratio $r_{52}=L'_{\rm CO(5-4)}/L'_{\rm CO(2-1)}$ for the nine gas-rich \squiggle galaxies with CO(5--4) and CO(2--1) observations \citep{bezansonNowYouSee2022,settonSQuIGGBuriedStar2025c,donofrioCOExcitation2026}. We find no clear relationship between $L_{1.4\,{\rm GHz}}$ and $r_{52}$, although the small sample and limited dynamic range preclude a strong statistical constraint. This comparison is presented separately because the CO(5--4) targets constitute a gas-rich subset rather than a representative sample of all 52 VLA targets. Moreover, elevated CO excitation is not unique to AGN heating and may also reflect processes such as compact star formation and shocks \citep[e.g.,][]{papadopoulosMolecularGasLuminous2012b,narayananTheoryExcitationCO2014a,kamenetzkyRecoveringPhysicalProperties2018b,frenchStateMolecularGas2023c}.

\clearpage
\bibliographystyle{aasjournal}
\bibliography{references.bib}

\end{CJK*}
\end{document}